%% file: ms.tex
\documentclass[11pt, a4paper]{article}

\title{Fractional homomorphism, Weisfeiler-Leman invariance,\\ and the Sherali-Adams hierarchy\\ for the Constraint Satisfaction Problem\thanks{An extended abstract of this work
appeared in \emph{Proceedings of the 46th International Symposium on Mathematical Foundations of Computer Science (MFCS 2021)}. The project that gave rise to these results received the support of a fellowship from ``la Caixa'' Foundation (ID 100010434). The fellowship code is LCF/BQ/DI18/11660056. This project has received funding from the European Union’s Horizon 2020 research and innovation programme under the Marie Skłodowska-Curie grant agreement No. 713673. Victor Dalmau was supported by MICCIN grants TIN2016-76573-C2-1P and PID2019-109137GB-C22.}}

\author{
Silvia Butti\\
Department of Information and Communication Technologies\\
Universitat Pompeu Fabra, Spain\\
\texttt{silvia.butti@upf.edu}
\and
V{\'{\i}}ctor Dalmau\\
Department of Information and Communication Technologies\\
Universitat Pompeu Fabra, Spain\\
\texttt{victor.dalmau@upf.edu}
}

\date{}

\usepackage[margin=1in]{geometry}
\usepackage{amsmath}
\usepackage{amsthm} 
\usepackage{amssymb}
\usepackage{hyperref}
\hypersetup{colorlinks=true,citecolor=blue,linkcolor=blue,urlcolor=blue}
\usepackage{amsfonts}
\usepackage{mathtools}
\usepackage{enumitem}
\usepackage{thmtools,thm-restate}
\usepackage[title]{appendix}
\allowdisplaybreaks

\newtheorem{theorem}{Theorem}
\newtheorem{corollary}[theorem]{Corollary}

\newtheorem{lemma}[theorem]{Lemma}

\theoremstyle{definition}

\theoremstyle{remark}
\newtheorem{claim}{Claim}

\newenvironment{claimproof}[1]{\par\noindent\textit{Proof of Claim}\space#1.}{\hfill $\blacksquare$}

\newcommand{\mb}[1]{\mathbf{#1}}
\newcommand{\csp}{\textnormal{CSP}}
\newcommand{\blp}{\textnormal{BLP}}

\newcommand{\bx}{\textbf{x}}
\newcommand{\by}{\textbf{y}}
\newcommand{\bb}{\textbf{b}}

\newcommand{\bd}{\textbf{d}}
\newcommand{\bc}{\textbf{c}}
\newcommand{\bz}{\textbf{z}}
\newcommand{\bt}{\textbf{t}}
\newcommand{\bq}{\textbf{q}}
\newcommand{\ba}{\operatorname{{\bf a}}}

\newcommand{\bA}{\mathbf{A}}
\newcommand{\bB}{\mathbf{B}}
\newcommand{\bE}{\mathbf{E}}
\newcommand{\bX}{\mathbf{X}}
\newcommand{\bC}{\mathbf{C}}
\newcommand{\bD}{\mathbf{D}}
\newcommand{\bP}{\mathbf{P}}
\newcommand{\bQ}{\mathbf{Q}}
\newcommand{\bT}{\mathbf{T}}

\newcommand{\bfa}{\mathbf{A}}
\newcommand{\bfb}{\mathbf{B}}
\newcommand{\bfx}{\mathbf{X}}
\newcommand{\xone}{\mathbf{X}_{1}}
\newcommand{\xtwo}{\mathbf{X}_{2}}

\newcommand{\stark}{*_{k}}
\newcommand{\astarkbf}{\mathbf{A}^\stark}
\newcommand{\bstarkbf}{\mathbf{B}^\stark}
\newcommand{\cstarkbf}{\mathbf{C}^\stark}
\newcommand{\dstarkbf}{\mathbf{D}^\stark}
\newcommand{\astark}{A^\stark}
\newcommand{\bstark}{B^\stark}

\newcommand{\Hom}{\textnormal{Hom}}
\newcommand{\sa}{\textnormal{SA}}
\newcommand{\sak}{\textnormal{SA}^k(\bA,\bB)}
\newcommand{\saone}{\textnormal{SA}^1(\bA,\bB)}
\newcommand{\sastar}{\textnormal{SA}^1(\astarkbf,\bstarkbf)}
\newcommand{\tjii}{T_{j,\textbf{i}}}

\newcommand{\rii}{R_{\textbf{i}}}

\newcommand{\pii}{\pi_{\textbf{i}}}
\newcommand{\pr}{\pi}
\newcommand{\RA}{R(\bfa)}
\newcommand{\RB}{R(\bfb)}
\newcommand{\arity}{\operatorname{arity}}
\newcommand{\na}{N_{\bA}}
\newcommand{\nb}{N_{\bB}}

\newcommand{\eqwl}[1][1]{\equiv_{#1}}

\begin{document}

\maketitle

\begin{abstract}
Given a pair of graphs $\bA$ and $\bB$, the problems of deciding whether there exists either a homomorphism or an isomorphism from $\bA$ to $\bB$ have received a lot of attention. While graph homomorphism is known to be NP-complete, the complexity of the graph isomorphism problem is not fully understood. A well-known combinatorial heuristic for graph isomorphism is the Weisfeiler-Leman test together with its higher order variants. On the other hand, both problems can be reformulated as integer programs and various LP methods can be applied to obtain high-quality relaxations that can still be solved efficiently. We study so-called fractional relaxations of these programs in the more general context where $\bA$ and $\bB$ are not graphs but arbitrary relational structures. We give a combinatorial characterization of the Sherali-Adams hierarchy applied to the homomorphism problem in terms of fractional isomorphism. Collaterally, we also extend a number of known results from graph theory to give a characterization of the notion of fractional isomorphism for relational structures in terms of the Weisfeiler-Leman test, equitable partitions, and counting homomorphisms from trees. As a result, we obtain a description of the families of CSPs that are closed under Weisfeiler-Leman invariance in terms of their polymorphisms as well as decidability by the first level of the Sherali-Adams hierarchy.
\end{abstract}

\input{0Intro-arxiv}
\input{1Definitions-arxiv}
\input{2Proof-arxiv}
\input{3Applicatons-arxiv}
\bibliography{biblio}
\input{Appendix-arxiv}

\end{document}

%% file: 0Intro-arxiv.tex
\section{Introduction}

The graph isomorphism and homomorphism problems, that is, the problems of deciding, given input graphs $\bA$ and  $\bB$, whether $\bA$
is isomorphic (respectively homomorphic) to $\bB$, have been the subject of extensive research. 
Despite an important research effort, it is still an open problem to determine whether the isomorphism problem for graphs can be solved in polynomial time. The recent quasipolynomial algorithm for the problem presented by Babai \cite{Babai2015} is widely regarded as a major breakthrough in theoretical computer science.

In turn, the complexity of the graph homomorphism problem has also been intensively studied in combinatorics (see \cite{HellNesetril:book}) as well as in a more general setting, known as the Constraint Satisfaction Problem (CSP),  where $\bA$ and $\bB$ are not required to be graphs but can be arbitrary relational structures. The CSP 
is general enough to encompass problems from areas as diverse as artificial intelligence, optimization, computer algebra, computational biology, computational linguistics, among many other. In contrast with the graph isomorphism problem it was quickly established that the homomorphism problem is NP-complete even for graphs, as it can encode graph coloring.
Consequently, an important research effort has been put into identifying tractable fragments of the problem, in particular by fixing the target structure $\bB$. This culminated in the recent major result of Bulatov \cite{Bulatov2017} and Zhuk \cite{Zhuk2017}, which confirmed a conjecture by Feder and Vardi \cite{feder1998computational} that, given a fixed target structure $\bB$, the homomorphism problem for $\bB$ is either solvable in polynomial time or NP-complete.

Linear programming relaxations, among other relaxations such as SDP-based, have been largely used in the study of both the isomorphism and the homomorphism problem. In fact, the isomorphism problem for graphs $\bA$, $\bB$ can be reformulated as an integer program which asks whether there exists a permutation matrix $X$ such that $X \na=\nb X$, where $\na$ and $\nb$ are the adjacency matrices of $\bA$, $\bB$ respectively. If we relax this condition to only require that $X$ is doubly stochastic, we obtain what is known as fractional isomorphism.

Fractional isomorphism has a combinatorial counterpart in the 1-dimensional Weisfeiler-Leman (1-WL) algorithm \cite{leman1968reduction}, also known as colour refinement. In particular, it was shown in \cite{Tinhofer1986, Tinhofer1991} and \cite{Ramana1994} that $\bA$ and $\bB$ are fractionally isomorphic if and only if 1-WL does not distinguish between them.
1-WL produces a sequence of colourings $c_0,c_1,\ldots$ of the nodes of a graph by means of an iterative refinement procedure which assigns a pair of nodes to the same colour class of $c^i$ if they belong to the same class of $c^{i-1}$, and additionally they have the same number of neighbours of each colour in $c^{i-1}$. The algorithm keeps iterating until a fixed point is reached. The Weisfeiler-Leman algorithm is a very powerful heuristic to test for graph isomorphism: if two graphs are distinguished by 1-WL (that is, they give rise to distinct fixed-point colourings up to renaming of the vertices), then this is a witness that the graphs are not isomorphic. In fact, it was shown that 1-WL decides the isomorphism problem on almost all graphs (that is, all but $o(2^{\binom{n}{2}})$ graphs on $n$ vertices for every $n$) \cite{Babai1980}. 

However, it is also easy to see that 1-WL fails on some very simple instances, such as regular graphs. 
To address these limitations, the original Weisfeiler-Leman algorithm has been extended so that at every iteration on a graph $\bA$ it produces a colouring of the set of $k$-tuples ($k > 1$) of nodes of $\bA$. 
It was initially conjectured that this hierarchy of increasingly powerful methods, known as the $k$-dimensional Weisfeiler-Leman ($k$-WL) algorithm, would provide a polynomial time graph isomorphism test at least for graphs of bounded degree. While this conjecture was proved to be incorrect \cite{Cai92}, $k$-WL turns out to have a number of useful applications, such as the important role it plays in the aforementioned quasipolynomial algorithm of Babai. 

In addition, the $k$-WL algorithm has proven to be very robust and has a number of equivalent formulations. In \cite{Dell2018}, a new characterization was given in terms of counting homomorphisms: for every $k \geq 1$, $\bA$ and $\bB$ are indistinguishable by the $k$-WL algorithm if for every graph $\bT$ of treewidth at most $k$, the number of homomorphisms from $\bT$ to $\bA$ is equal to the number of homomorphisms from $\bT$ to $\bB$. Further, it has been shown in \cite{Cai92}
that $\bA$ and $\bB$ are indistinguishable by the $k$-WL algorithm if and only if they are indistinguishable in the logic $C^{k+1}$ consisting of all FO formulas with counting quantifiers that have at most $k+1$ variables.

Similarly to the Weisfeiler-Leman hierarchy for fractional isomorphism, any LP relaxation of an integer $0/1$-program, like
that of graph isomorphism, can be further strengthened by sequentially applying so-called lift-and-project methods from mathematical programming in order to obtain a hierarchy of increasingly tighter relaxations which can still be solved efficiently. The main idea behind these is to add auxiliary variables and valid inequalities to an initial relaxation of a $0/1$ integer program.
These methods, which include Lov\'asz-Schrijver \cite{Lovasz1991} and Sherali-Adams \cite{Sherali1990}, have been used to study classical problems in combinatorial optimization such as Max-Cut, Vertex Cover, Maximal Matching, among many others. 

Quite surprisingly, Atserias and Maneva \cite{Atserias2013} and Malkin \cite{Malkin2014} were able to lift the connection between the 1-WL algorithm and fractional isomorphism to show a close correspondence between the higher levels of the SA hierarchy for the graph isomorphism problem and the $k$-WL algorithm (and, hence, with the logic $C^{k+1}$). This correspondence was further tightened in \cite{Grohe2015Pebble}.

Let us now turn our attention to the homomorphism problem. Here, LP relaxations have been intensively used in the more general setting of the Constraint Satisfaction Problem and, most usually, in the approximation of its optimization versions such as MaxCSP, consisting of finding a map  from the universe of $\bA$ to the universe of $\bB$ that maximizes the number of constraints satisfied, among other variants. For instance, a simple linear programming relaxation yields a $2$-approximation algorithm for the Vertex Cover problem, and no better polynomial time approximation algorithm is known.

One of the simplest and most widespread LP relaxations of the homomorphism problem 
(see for example \cite{Kun2012}) has an algebraic characterization that very much resembles that of fractional isomorphism. For the sake of simplicity we shall present it here in the restricted case of graphs. 
Let us start with the algebraic formulation of fractional isomorphism which can be, alternatively, expressed as the existence of a pair of doubly stochastic matrices $X$ and $Y$ such that $X M_{\bA}=M_{\bB} Y$ and $M_{\bA} Y^T=X^T M_{\bB}$, where $M_{\bA}$ denotes the {\em incidence}  matrix of $\bA$. If we relax this condition to only require that $X$ and $Y$ are left stochastic and, additionally, we drop the second equation, then we obtain a relaxation of graph homomorphism, which we call fractional homomorphism. With some minor variations depending on whether the objective function is present (as in MaxCSP) or not and how repeated elements in a tuple are treated, this LP formulation has been extensively used \cite{BrakensiekGWZ20,DalmauK13,DalmauKM18,GhoshT18,KumarMTV11,Kun2012}.

We consider relaxations arising from the application of the Sherali-Adams (SA) method. Giving an explicit description of the inequalities produced by the SA method for homomorphism might be relatively cumbersome (because the constraints of the input structures $\bA$ and $\bB$ must be encoded in the polytope-defining inequalities rather than in the objective function as in the optimization variants) even when the target structure $\bB$ is fixed. Hence, we consider a simpler family of LP inequalities, interleaved with the SA hierarchy, of which fractional homomorphism corresponds to the first level.
Our hierarchy coincides with the usual SA hierarchy for MaxCSP \cite{ChanLRS13,GeorgiouMT09,ThapperZ17,YoshidaZ14} where the objective function has been turned into a constraint.

Our main result is a combinatorial characterization of this family of LP relaxations which, abusing slightly notation, we still shall refer to as $\sa^k,k\geq 1$. 
Along the way, we extend a number of the aforementioned results from graph isomorphism to general isomorphism between relational structures which yields a hierarchy of relaxations of (relational structure) isomorphism. Our results show that, for every $k\geq 1$, the $k$th level relaxation of the homomorphism problem, that is, $\sa^k$, is tightly related with the corresponding 
$k$th level relaxation of isomorphism, which we denote using $\eqwl[k]$. 

Let us consider the first level of the hierarchy, i.e, fractional homomorphism and fractional isomorphism. In the case of graphs $\eqwl[1]$-equivalence coincides precisely with $1$-WL equivalence. Here we show that  a structure $\bA$ is fractionally homomorphic to a structure $\bB$ if and only if there exists a sequence of structures $\bX_0,\bX_1,\dots,\bX_n$ with $\bA=\bX_0$, $\bB=\bX_n$, and for every $i<n$, $\bX_i$ is either homomorphic or $\eqwl[1]$-equivalent to $\bX_{i+1}$. The correspondence for any higher level $k$ is obtained by replacing 
$\bX_0$ and $\bX_n$ by suitably defined structures $\astarkbf$, $\bstarkbf$ that allow to reduce the $k^{th}$ level of the hierarchy to the first level.

In particular, it follows that feasibility in $\sa^k$ is preserved by structures that are $\eqwl[k]$-equivalent.

In essence, this is due 
to the fact that the LP relaxation of homomorphism inherits the symmetries of $\bA$ and $\bB$. There
are, indeed, 
several results exploring the symmetries of an LP program in a similar fashion. In \cite{Atserias2013} such types of symmetries are used to transfer results between the logic $C^k$ and LP relaxations of several combinatorial problems, whereas \cite{GroheKMS14} aims to identify the partition classes (as in the WL algorithm) of the variables and constraints of an LP program so that, by identifying those in the same class, the LP size is reduced. That being said, the main interest of our result lies precisely in the opposite direction: that is, the fact that the fractional homomorphism LP relaxation is able to certify that $\bA$ is not homomorphic to $\bB$ unless $\bA$ belongs to the backwards closure of $\bB$ under homomorphism and $\eqwl[1]$ equivalence, or, even more strongly, unless $\bA$ is homomorphic to a structure $\bX_1$ which is $\eqwl[1]$-equivalent to a structure $\bX_2$ which in turn is homomorphic to $\bB$.

We apply our results to study the following question: for which structures $\bB$ is the set $\csp(\bB)$, which contains all the structures homomorphic to $\bB$, closed under 1-WL equivalence. This question arises in the context of solving CSPs in a distributed manner \cite{Butti2021} where the elements of the input instance (nodes, edges) are distributed among agents which communicate with each other by sending messages through fixed communication channels. Using our main results we can rederive the classification obtained in \cite{Butti2021} using substantially different techniques.

%% file: 1Definitions-arxiv.tex
\section{Preliminaries}

\paragraph*{Relational structures.}

For a positive integer $n$, We denote by $[n]$ the set $\{1,\ldots,n\}$.
We shall denote tuples in boldface. Let $\textbf{a}=(a_1,\dots,a_k) \in A^k$. We use $\textbf{a}[i]$ to denote $a_i$ and $\{\textbf{a}\}$ to denote the set of variables which occur in $\textbf{a}$. For every tuple ${\bf i}=(i_1,\dots,i_n)\in [k]^n$ we use $\pr_{\bf i} \ba$ to denote the {\em projection} of $\ba$ to $\bf i$, i.e, the tuple 
$(a_{i_1},\dots,a_{i_n})$. If $I\subseteq [k]$ we might abuse slightly notation and use $\pr_I \ba$ to
refer to $\pr_{\bf i} \ba$ where $\bf i$ is the tuple that contains the elements of $I$ in increasing order.
For every function $f$ on a domain containing $\{\textbf{a}\}$, we denote by $f(\textbf{a})$ the coordinate-wise application of $f$ to $\textbf{a}$.

Given a set $A$ and a positive integer $k$, a $k$-ary \textit{relation} over $A$ is a subset of $A^k$. A \textit{signature} $\sigma$ is a finite collection of relation symbols, each with an associated arity. We shall use $\arity(R)$ to denote the arity of a relation symbol $R$. A \textit{relational structure} $\bfa$ over $\sigma$, or simply a $\sigma$-structure, consists of a set $A$ called the \textit{universe} of $\bfa$, and a relation $\RA$ over $A$ for each $R \in \sigma$ of the corresponding arity. We denote by $\mathcal{C}_\bfa$ the set $\{(\mb{a},R) \mid \mb{a} \in \RA, R \in \sigma\}$. Elements $(\mb{a},R)$ of $\mathcal{C}_\bfa$ will alternatively be denoted $R(\textbf{a})$ and will be referred to as \textit{constraints}. We shall usually use the same boldface and (standard) capital letter to refer to a structure and its universe, respectively.
A \textit{graph} is a relational structure whose signature consists of a single binary relation that is symmetric and non-reflexive.

Let $\bfa$, $\bfb$ be $\sigma$-structures. A \textit{homomorphism} from $\bfa$ to $\bfb$ is a map $h: A \to B$ such that for every $R \in \sigma$ and every $\mb{a} \in \RA$ it holds that $h(\textbf{a}) \in \RB$.
If there exists a homomorphism from $\bfa$ to $\bfb$ we say that $\bfa$ is homomorphic to $\bfb$ and we write $\bfa \to \bfb$.
We shall use $\Hom(\bA;\bB)$ to denote the number of homomorphisms from $\bA$ to $\bB$. The problem of deciding, given two similar structures $\bA$ and $\bB$, whether $\bA$ is homomorphic to $\bB$ is known as
the \textit{Constraint Satisfaction Problem} (CSP). If we fix the target structure $\bB$ so that the input is only $\bA$, then we obtain the  Constraint Satisfaction Problem  over $\bfb$, denoted $\csp(\bfb)$.

An \textit{isomorphism} from $\bfa$ to $\bfb$ is a bijective map $f: A \to B$ such that for every $R \in \sigma$  and every $\textbf{a} \in A^{\arity(R)}$  it holds that  $\mb{a} \in \RA$ if and only if $f(\textbf{a}) \in \RB$. 

The {\em union} $\bA\cup\bB$ of two $\sigma$-structures $\bA$ and $\bB$ is the structure $\bC$
with $C=A\cup B$ and $R(\bC)=R(\bA)\cup R(\bB)$ for every $R\in\sigma$. The {\em disjoint union} of two structures
$\bA$ and $\bB$ is the structure $\bA\cup \bC$ where $\bC$ is any $\sigma$-structure
isomorphic to $\bB$ satisfying $A\cap C=\emptyset$. We say that a structure is {\em connected} if it cannot be expressed as the disjoint union of two structures. We say that $\bA$ is a {\em substructure} of $\bB$ if $\bA\cup\bB=\bB$. If, in addition,
$R(\bA)=R(\bB)\cap A^{\arity(R)}$ for every $R\in\sigma$ then $\bA$ is the substructure
of $\bB$ {\em induced} by $A$.

\paragraph*{Tree-like structures.} 

We define the \textit{factor graph}\footnote{We note that the definition of factor graph presented here differs from that in \cite{Butti2021}, in which the edges are labelled. We also note that the notion of factor graph, although similar, differs in several ways from the incidence multigraph (see \cite{LaroseLT07}) as the latter allows for parallel edges.} of a structure $\bA$ to be the bipartite graph with nodes $A \cup \mathcal{C}_{\bA}$ and where every $R(\ba)$ in $\mathcal{C}_{\bA}$ is joined by an edge with every $a\in\{\ba\}$. 

Then, we say that a structure $\bT$ is an \textit{ftree} (factor tree) if its factor graph is a tree in the ordinary graph-theoretic sense. If $\bQ$ is a substructure of $\bT$ and $\bQ$ is an ftree then we say that $\bQ$ is a \textit{subftree} of $\bT$.

A {\em tree-decomposition} of a structure $\bA$ is a pair $(G,\beta)$ where $G=(V,E)$ is a tree and $\beta:V\rightarrow{\mathcal P}(A)$ is a mapping such that the following conditions are satisfied:
\begin{enumerate}
    \item For every constraint $R(\ba)$ in $\mathcal{C}_{\bA}$ there exists a node $v\in V$ such that $\{\ba\}\subseteq\beta(v)$
    \item If $a\in\beta(u)\cap \beta(v)$ then $a\in\beta(w)$ for every node $w$ in the unique path in $G$ joining $u$ to $v$.
    \end{enumerate}
The {\em width} of a tree-decomposition $(G,\beta)$ is $\max\{|\beta(v)| \mid v\in V\}-1$ and the \textit{treewidth} of $\bA$ is defined as the smallest width among all its tree-decompositions.

\paragraph*{The Sherali–Adams hierarchy.}

In this presentation we follow \cite{Atserias2013}. Let $P\subseteq[0,1]^n$ be a polytope $\{\bx\in \mathbb{R}^n : M \bx\geq \bb, 0\leq\bx\leq 1\}$ for a matrix $M\in\mathbb{R}^{m\times n}$, and a column vector $\bb\in \mathbb{R}^m$. We denote the convex hull of the $\{0,1\}$-vectors in $P$ by $P^{\mathbb{Z}}$. The sequence of Sherali-Adams relaxations of $P^{\mathbb{Z}}$ is the sequence of polytopes $P=P^1\supseteq P^2\supseteq\cdots$ where $P^k$ is defined in the following way.

Each inequality in $M \bx\geq\bb$ is multiplied by all possible terms of the form $\Pi_{i\in I} x_i\Pi_{j\in J} (1-x_j)$ where $I,J\subseteq[n]$ satisfy $|I\cup J|\leq k-1$ and $I\cap J=\emptyset$. This leaves a system of polynomial inequalities, each of degree at most $k$. Then, this system is {\em linearized} and hence relaxed in the following way: each square $x_i^2$ is replaced by $x_i$ and each resulting monomial $\Pi_{i\in K} x_i$ is replaced by a variable $y_K$. In this way we obtain a polytope $P^k_L$. Finally, $P^k_L$ is projected back to $n$ dimensions by defining 
\[P^k:=\{\bx\in \mathbb{R}^n: \text{there exists $\by\in P^k_L$ such that $\by_{\{i\}}=\bx_i$ for each $i\in [n]$}\}.\]
We note here  that $P^{\mathbb{Z}}\subseteq P^k$ for every $k\geq 1$.

In order to apply the SA method to the homomorphism problem there are different possible choices for the polytope $P$ (encoding a relaxation of homomorphism) to start with, each one then yielding a different hierarchy.

Here we shall adapt a SA-based family of relaxations commonly used in optimization variants of CSP \cite{ChanLRS13,GeorgiouMT09,ThapperZ17,YoshidaZ14} which we  transform into a relaxation of (plain) CSP by just turning the objective function into a set of new restrictions. Hence, the resulting system of inequalities is not, strictly speaking, obtained using the SA method. Nonetheless, we shall abuse slightly notation and still use $\sa^k$ to refer to our system of inequalities. 

In fact, giving an explicit description of all inequalities obtained using the SA method for any natural polytope $P$ encoding the LP relaxation for a {\em general} CSP in our setting is a bit cumbersome (because the constraints of the CSP are encoded in the polytope-defining inequalities instead of the objective function as in CSP optimization variants). Hence, it seems sensible to settle for a good approximation as $\sa^k$. Indeed, as it can be seen in Appendix \ref{appendix:SA},
the sequence of relaxations $\sa^k$ is tightly interleaved with the sequence $P^k$ obtained by the SA method, in stricto sensu, for a natural choice of initial polytope $P$.

Given two structures $\bfa$ and $\bfb$, the system of inequality $\sak$ for the homomorphism problem over $(\bfa,\bfb)$ contains a variable $p_V(f)$ for every $V \subseteq A$ with $1 \leq |V| \leq k$ and every $f:V \to B$, and a variable $p_{R(\textbf{a})}(f)$ for every $R(\textbf{a}) \in \mathcal{C}_\bfa$ and every $f:\{\textbf{a}\} \to B$. Each variable must take a value in the range $[0,1]$. The variables are constrained by the following conditions:
\begin{align}
&\sum_{f:V \to B} p_V(f)=1  &&  V\subseteq A \textnormal{ s.t. } |V|\leq k  \label{eq:SA1} \tag{\sa1} \\
&p_U (f) = \sum_{\mathclap{g:V \to B,g|_{U} =f}} \, p_V(g) && U \subseteq V \subseteq A \, \textnormal{ s.t. }|V|\leq k, f:U \to B \label{eq:SA2} \tag{\sa2} \\
&p_U(f) = \sum_{\mathclap{g:V\to B, g|_U =f}} \, p_{R(\textbf{a})}(g) &&  R(\ba)\in{\mathcal C}_{\bA}, U\subseteq\{\ba\}=V \textnormal{ s.t. } |U| \leq k, f: U \to B \label{eq:SA3}\tag{\sa3} \\
&p_{R(\textbf{a})}(f) = 0 &&  R(\ba) \in \mathcal{C}_\bfa,  f : \{\textbf{a}\} \to B  \textnormal{ s.t. }f(\textbf{a}) \not \in \RB \label{eq:SA4}\tag{\sa4}
\end{align}

For the particular case of $k=1$ we shall use the simplified notation $p_v(f(v))$ to denote the variable $p_V(f)$ for a singleton set $V=\{v\}$ and a function $f:V \to B$.

\paragraph*{The transformation $\stark$.}

For every $k>0$ we define an operator $*_k$ that maps a structure into a new structure. As we shall see later this operator will allow to reduce $\sa^k$ feasibility to $\sa^1$ feasibility.

Let $\bfa$ be a $\sigma$-structure. Then we define the universe of $\astarkbf$ to be $\astark := \cup_{j \leq k} A^j \cup \mathcal{C}_{\bA}.$
Additionally, $\astarkbf$ contains the following relations:
\begin{spreadlines}{7pt}
\begin{align*}
T_{j,S}(\astarkbf)&=\{ (a_1,\dots,a_j)\in A^j \mid a_i=a_{i'} \: \forall i,i'\in S\}  &&  j\leq k, \, S\subseteq[j]\\
\tjii(\astarkbf)\,\,&=\{(\textbf{a},\pii\textbf{a}) \mid \textbf{a} \in A^j\}  &&  j',j\leq k, \, \textbf{i}\in [j]^{j'}\\
R_S(\astarkbf)\,\,\,&=\{ (a_1,\dots,a_{\arity(R)})\in R(\bA) \mid a_i=a_{i'} \: \forall i,i'\in S\} &&  R\in\sigma, \, S\subseteq[\arity(R)]\\
R_{\bf i}(\astarkbf)\,\,\,\,&=\{ (R(\ba),\pii\ba) \mid \ba\in R(\bA)\} &&  R\in\sigma, \, j\leq k, \, \textbf{i}\in [\arity(R)]^{j}.
\end{align*}
\end{spreadlines}
Then we have:

\begin{restatable}{lemma}{lestarkSherali} \label{le:starkSherali}
Let $\bfa$, $\bfb$ be $\sigma$-structures. Then $\sak$ is feasible if and only if $\sastar$ is feasible.
\end{restatable}

\paragraph*{Iterated degree, fractional isomorphism, and equitable partitions.}

In order to prove our main result we need to lift the known equivalence between fractional isomorphism and the WL algorithm from graphs to relational structures. 
 
Let $L := \{(S,R) \mid R \in \sigma, S \subseteq [\arity(R)] \}$ be a set, the elements of which we shall
call {\em labels}. We construct the \textit{matrix representation} $M_{\bfa}$ of a $\sigma$-structure $\bfa$ as follows (it will be convenient to assume that the indices of the rows and columns of a matrix are arbitrary sets). $M_{\bfa}$ is an $A \times \mathcal{C}_{\bfa}$ matrix whose entries are elements of $L$. In particular, for all $a \in A$ and $R(\textbf{a}) \in \mathcal{C}_\bfa$, we have that $ M_{\bfa}[a,R(\textbf{a})] = (S,R) $ where $S$ is the set containing all elements $i\in[\arity(R)]$ such that $a=\textbf{a}[i]$. 

In a nutshell we are lifting from graph isomorphism to matrix isomorphism, where two matrices are isomorphic if they are identical modulo a permutation of the rows and columns. To formalize this, it will be convenient to associate $\bfa$ with a set of 0-1 incidence matrices. In particular, for every $\ell \in L$ we define $M_{\bfa}^{\ell} \in \{0,1\}^{A \times \mathcal{C}_\bfa}$ as follows: $M_{\bfa}^{\ell}[a,R(\textbf{a})]=1$ if $ M_{\bfa}[a,R(\textbf{a})]=\ell$ and $M_{\bfa}^{\ell}[a,R(\textbf{a})]=0$ otherwise. In \cite{GroheKMS14}, relaxations of matrix isomorphism are also considered although in that setting the matrices have real entries and the goal is different from ours. 

We now describe a procedure akin to the 1-dimensional Weisfeiler-Leman algorithm to calculate iterative refinements of a colouring of the universe and constraint set of a relational structure. While there are syntactical differences, when run on graphs this procedure is equivalent for all purposes to 1-WL.
For every $k\geq 0$ and $x\in A\cup \mathcal{C}_\bfa$, we define inductively the \textit{iterated degree} $\delta^{\bfa}_k(x)$ of $x$ on $\bfa$ as follows.
We set $\delta_{0}^\bfa(x)$ to be one of two arbitrary symbols that distinguish elements of $A$ from  elements of $\mathcal{C}_\bfa$. For $k\geq 1$ we set $\delta_k^\bfa(a)=\{\{(\ell,\delta_{k-1}^\bfa(\textbf{a},R)) \mid M_{\bfa}^{\ell}[a, R(\textbf{a})]=1\}\}$ and $\delta_k^\bfa(\textbf{a},R)=\{\{(\ell,\delta_{k-1}^\bfa(a)) \mid M_{\bfa}^{\ell}[a, R(\textbf{a})]=1\}\}$, where double curly brackets denote that $\delta_k^\bfa(x)$ is a multiset. We say that $\bfa$ and $\bfb$ have the same iterated degree sequence 
if for every $k\geq 0$, $\{\{\delta^\bfa_k(a) \mid a\in A\cup \mathcal{C}_{\bA}\}\}=\{\{\delta^\bfb_k(b) \mid b\in B\cup \mathcal{C}_{\bB}\}\}$. 
Note that if there exists a matrix isomorphism from $\bfa$ to $\bfb$ (or alternatively, $\bA$ and $\bB$ are isomorphic) then $\bfa$ and $\bfb$ have the same iterated degree sequence, but the converse does not hold.

The notion of equitable partition is key in the proof of the equivalence of the different characterizations of fractional isomorphism. We present its adaptation to relational structures.
A \textit{partition} of a $\sigma$-structure $\bfa$ is a pair $(P,Q)$ where $P=\{P_i \mid i\in I\}$ is a partition of $A$ and $Q=\{Q_j \mid j\in J\}$ is a partition of $\mathcal{C}_\bfa$. We say that
$(P,Q)$ is \textit{equitable} if for every $i\in I$, $j\in J$, and $\ell\in L$, there are integers $c^{\ell}_{i,j},
d^{\ell}_{j,i}$, called the {\em parameters} of the partition, such that for every every $i\in I$, every $a\in P_i$, every $\ell\in L$, and every $j\in J$, we have \begin{equation} \label{eq:eqpartQ} \tag{P1}
|\{(\textbf{a},R)\in Q_j \mid M_{\bfa}[a,R(\ba)]=\ell\}|=c^{\ell}_{i,j}
\end{equation}
\noindent and, similarly, for every $j\in J$, every $R(\ba)\in Q_j$,  every $\ell\in L$, and every $i\in I$  we have \begin{equation} \label{eq:eqpartP} \tag{P2}
|\{a\in P_i \mid M_{\bfa}[a,(\textbf{a},R)]=\ell\}|=d^{\ell}_{j,i}. \end{equation} 
We say that two structures $\bA$, $\bB$ have a {\em common equitable partition} if there are equitable partitions 
$(\{P_i^\bA \mid i\in I\},\{Q_j^\bA \mid j\in J\})$ and $(\{P_i^\bB \mid i\in I\},\{Q_j^\bB \mid j\in J\})$ of $\bA$
and $\bB$ with the same parameters satisfying $|P^\bA_i|=|P^\bB_i|$ for every $i\in I$ and $|Q^\bA_j|=|Q^\bB_j|$ for every $j\in J$.
We note that if $\bA$ and $\bB$ are connected it is not necessary to verify this latter requirement and, instead, it is enough to check that $|A|=|B|$.

A matrix $M$ of non-negative real numbers is said to be \textit{left} (resp. \textit{right}) \textit{stochastic} if all its columns (resp. rows) sum to $1$. Note that we do not require $M$ to be square. A \textit{doubly stochastic matrix} is a square matrix that is both left and right stochastic.

\begin{restatable}{theorem}{ledigraph}
\label{le:digraph}
Let $\bfa$, $\bfb$ be $\sigma$-structures. The following are equivalent:
\begin{enumerate} 
\item There exist doubly stochastic matrices $X$, $Y$ such that $XM_{\bfa}^{\ell}=M_{\bfb}^{\ell}Y$ and $M_{\bfa}^{\ell}Y^T=X^TM_{\bfb}^{\ell}$ for every $\ell\in L$; \label{1digraphItem}
\item $\bfa$ and $\bfb$ have the same iterated degree sequence; \label{2digraphItem}
\item $\bA$ and $\bB$ have a common equitable partition; \label{3digraphItem}
\item $\Hom(\bT,\bA)=\Hom(\bT,\bB)$ for all $\sigma$-ftrees $\bT$. \label{4digraphItem}
\end{enumerate}
If, additionally, $\bfa$ and $\bfb$ are graphs, then the following is also equivalent:
\begin{enumerate}
    \setcounter{enumi}{4}
    \item There exists a doubly stochastic matrix $X$ such that $X\na = \nb X$ where
    $\na$ and $\nb$ denote the adjacency matrices of $\bfa$ and $\bfb$ respectively. \label{5digraphItem}
\end{enumerate}
\end{restatable}

The equivalence between (\ref{1digraphItem}), (\ref{2digraphItem}) and (\ref{3digraphItem}) is an immediate generalization of the equivalence between the algebraic and combinatorial characterizations of fractional graph isomorphism (see for example \cite{Scheinerman2011fractional}), while (\ref{4digraphItem}) generalizes the corresponding characterization of graphs from \cite{Dell2018} in terms of counting homomorphisms from trees. Also, note that condition (\ref{5digraphItem}) shows that our definition coincides with the standard notion of fractional isomorphism when $\bfa$ and $\bfb$ are graphs. 

Similarly to the case of graphs, the notion of fractional isomorphism captured in Theorem \ref{le:digraph} can be strengthened giving rise to a hierarchy of increasingly tighter relaxations. In particular, for every $k\geq 1$, we shall denote $\bA \eqwl[k] \bB$ whenever $\astarkbf$ and $\bstarkbf$ satisfy the conditions of Theorem $\ref{le:digraph}$.

It is easy to see that the case $k=1$ would be unchanged if one replaces $\bA^*_1$ and $\bB^*_1$ by $\bA$ and $\bB$ respectively and, hence, Theorem \ref{le:digraph} characterizes $\eqwl[1]$. For other small values of $k$ other than $k=1$, $\eqwl[k]$ is a bit more difficult to characterize. However, as long as $k$ is at least as large as the arity of any relation in the signature then we have the following
result.

\begin{restatable}{lemma}{homTreewidth} \label{le:homTreewidth}
Let $r$ be the maximum arity among all relations in $\sigma$ and assume that $r\leq k$. Then for every pair of structures $\bfa$, $\bfb$ the following are equivalent:
\begin{enumerate}
    \item $\bfa \eqwl[k] \bB$ \label{1homTreewidthItem}
    \item $\Hom(\bQ;\bA)=\Hom(\bQ;\bB)$ for every structure $\bQ$ of treewidth $<k$. \label{2homTreewidthItem}
\end{enumerate}

\end{restatable}

This result, which follows easily from condition (\ref{4digraphItem}) in Theorem \ref{le:digraph}, is inspired by a similar result \cite{Dell2018} which states that two graphs $\bA,\bB$ are indistinguishable by the $k$-WL algorithm if and only if $\Hom(\bQ;\bA)=\Hom(\bQ;\bB)$ for every graph $\bQ$ of treewidth $\leq k$. It is not that surprising that a similar result can be shown for $\eqwl[k]$ since, after all, the $k$-WL algorithm can be seen as the $1$-WL algorithm applied to $k$-ary tuples. However, note
that the bound on the treewidth differs in one unit between $k$-WL and $\eqwl[k]$. Still, using Lemma \ref{le:homTreewidth}
it can be shown that for $r\leq k$, $\eqwl[k]$ can be alternatively characterized in logical terms extending, again, an analogous result for $k$-WL \cite{ImmermanL90}. More precisely, $\bA\eqwl[k]\bB$ if and only $\bA$ and $\bB$ satisfy the same formulas in the $k$-variable fragment of first-order logic with counting quantifiers, denoted $C^k$ (we omit the definition as it will not be needed).

\subsection{Main results}
The main result of this paper is a new characterization in terms of fractional isomorphism of the Sherali-Adams relaxation of the homomorphism problem.

\begin{theorem} \label{thm:SAk}
Let $\bfa$, $\bfb$ be relational structures. Then, the following are equivalent:
\begin{enumerate}
    \item $\sa^k(\bfa,\bfb)$ is feasible;
    \item There exists a sequence of structures $\bfx_0, \ldots, \bfx_n$ such that $\bfx_0=\astarkbf$, $\bfx_n=\bstarkbf$, and for all $i=0,\ldots,n-1$ we have that $\bfx_i \to \bfx_{i+1}$ or $\bfx_{i} \eqwl[1] \bfx_{i+1}$;
    \item There exist structures $\bfx_1, \bfx_2$ such that $\astarkbf \to \bfx_{1}$, $\bfx_{1} \eqwl[1] \bfx_{2}$, and $\bfx_2 \to \bstarkbf$. 
\end{enumerate}
\end{theorem}

Theorem \ref{thm:SAk} is an immediate consequence, using Lemma \ref{le:starkSherali}, of the following Theorem.

\begin{theorem}\label{thm:SA1}
Let $\bfa$, $\bfb$ be relational structures. Then, the following are equivalent:
\begin{enumerate}
    \item $\sa^1(\bfa,\bfb)$ is feasible; \label{1SA1Item}
    \item  There exist left stochastic matrices $X$, $Y$ such that for every $\ell=(S,R) \in L$, it holds that
    $XM_{\bfa}^{\ell}\leq \sum_{\ell'} M_{\bfb}^{\ell'}Y$ where $\ell'$ ranges over all $(S',R)\in L$ with $S\subseteq S'$; \label{2SA1Item}
    \item There exists a sequence of structures $\bfx_0, \ldots, \bfx_n$ such that $\bfx_0=\bfa$, $\bfx_n=\bfb$, and for all $i=0,\ldots,n-1$ we have that $\bfx_i \to \bfx_{i+1}$ or $\bfx_{i} \eqwl[1] \bfx_{i+1}$; \label{3SA1Item}
    \item There exist structures $\bfx_1, \bfx_2$ such that $\bfa \to \bfx_{1}$,  $\bfx_{1} \eqwl[1] \bfx_{2}$, and $\bfx_2 \to \bfb$. \label{4SA1Item}
\end{enumerate}
If in addition $\bA$ and $\bB$ have no loops (meaning that there are no repeated elements in any constraint) then the following condition is also equivalent:
\begin{enumerate}
    \setcounter{enumi}{4}
    \item  There exist left stochastic matrices $X$, $Y$ such that for every $\ell \in L$ it holds that 
    $XM_{\bfa}^{\ell}=M_{\bfb}^{\ell}Y$.\label{5SA1Item}
\end{enumerate}
\end{theorem}
Note that for graphs condition (\ref{5SA1Item}) of the above theorem is naturally seen as the homomorphism counterpart of the notion of fractional isomorphism (see condition (\ref{1digraphItem}) in Lemma \ref{le:digraph}). Consequently, we shall say that $\bA$ is fractionally homomorphic to $\bB$ whenever $\bA$ and $\bB$ satisfy the conditions of Theorem \ref{thm:SA1}.

While Theorems \ref{thm:SAk} and \ref{thm:SA1} have applications, for instance in the field of Constraint Satisfaction Problems (see Section \ref{sec:applications}), we believe that their main interest is that they shed light on the LP relaxations for homomorphism as they show that fractional homomorphism and its higher order counterparts can be decomposed into a sequence of basic, better studied morphisms, a fact which we find interesting on its own. Additionally, they establish a close link between LP relaxations for isomorphism and homomorphism, which were introduced initially in different fields.

%% file: 2Proof-arxiv.tex
\section{Proof of Theorem \ref{thm:SA1}}

The equivalence $(\ref{1SA1Item})\Leftrightarrow(\ref{2SA1Item})$ is merely syntactic. In particular we shall show that there is a one-to-one satisfiability-preserving correspondence between pairs of matrices and variable assignments of $\saone$. However, we first need to massage a bit the two formulations. First, we can assume that for every $R^A(\textbf{a})\in {\mathcal C}_{\bA}$ and $R^B(\textbf{b}) \in {\mathcal C}_{\bB}$, the corresponding entry in $Y$ is null unless $R^A=R^B$ and $f(\ba)=\bb$ for some  $f:\{\ba\}\rightarrow\{\bb\}$, since otherwise there is no way that $Y$ can be part of a feasible solution. Secondly, we note that the feasibility of $\saone$ does not change if in (\ref{eq:SA3}) we replace
$=$ by $\leq$ obtaining a new set of inequalities (which to avoid confusion we shall denote by (\hypertarget{eq:SA3prime}{$\sa3'$})) and, in addition, we add
for every $R(\bf a) \in \mathcal{C}_\bfa$ the equality
\begin{equation} \label{eq:SA5}\tag{\sa5}
\sum_{f:\{\ba\}\rightarrow B} p_{R(\ba)}(f)=1.
\end{equation}
Finally, note that in $\saone$ we can ignore (\ref{eq:SA2}).

Then we can establish the following correspondence between pairs of matrices $X$, $Y$ and assignments $\saone$:
for every $a\in A$ and $b\in B$, we set $p_a(b)=X[b,a]$ and for every $R(\mb{a}) \in \mathcal{C}_\bfa$ and $f:\{\textbf{a}\} \to B$  we define $p_{R(\mb{a})}(f)=Y[R(f(\mb{a})),R(\mb{a})]$. Then, it is easy to see that (\hyperlink{eq:SA3prime}{$\sa3'$}) corresponds to $XM_{\bfa}^{\ell}\leq \sum_{\ell' \in L_\ell} M_{\bfb}^{\ell'}Y$ for every $\ell\in L$ (where $L_{(S,R)}:=\{(S',R) \in L \mid S \subseteq S'\}$), $X$ being left stochastic corresponds to (\ref{eq:SA1}), and $Y$ being left stochastic corresponds to (\ref{eq:SA5}).

The equivalence $(\ref{1SA1Item}) \iff (\ref{5SA1Item})$ is obtained as in $(\ref{1SA1Item}) \iff (\ref{2SA1Item})$. We just need to notice that when $\bA$ and $\bB$ have no loops, then none of the entries in $M_{\bA}$ and $M_{\bB}$ contain any label $\ell=(S,R)\in L$ where $|S|>1$ and hence it is only necessary to consider labels $\ell=(S,R)\in L$
where $S$ is a singleton. Observe that, in this case, the equation in $(\ref{2SA1Item})$ becomes 
$XM_{\bfa}^{\ell}\leq M_{\bfb}^{\ell}Y$ since for every label $\ell=(S,R)$ where $S$ is a singleton, the only label $(S',R)$ with $S\subseteq S'$ and $M_{\bfb}^{\ell}$ not a zero matrix is $\ell$ itself. Finally, in order to 
replace $\leq$ by $=$ in the previous equation we just need to use (\ref{eq:SA3}) instead of (\hyperlink{eq:SA3prime}{$\sa3'$}).

Notice that $(\ref{4SA1Item}) \implies (\ref{3SA1Item})$ is trivial.

The proof of $(\ref{3SA1Item}) \implies (\ref{2SA1Item})$ is by induction on $n$. If $n=0$ the claim is immediate, so assume that $n \geq 1$. Let $\bX_0, \bX_1, \ldots, \bX_n$ be a sequence of structures satisfying (\ref{3SA1Item}). By the induction hypothesis, there exist left stochastic matrices $X$, $Y$ such that $XM_{\bfx_1}^{\ell}\leq \sum_{\ell' \in L_\ell} M_{\bfx_n}^{\ell'}Y$ for all $\ell\in L$.

If $\bfx_0 \eqwl[1] \bfx_1$ then it follows from Theorem \ref{le:digraph} that there exist doubly stochastic matrices $X'$ and $Y'$ such that $X'M_{\bfx_0}^{\ell}= M_{\bfx_1}^{\ell}Y'$ for all $\ell \in L$, and so it is easy to verify that $XX'$, $YY'$ are such that (\ref{2SA1Item}) holds. Assume that $\bfx_0 \to \bfx_1$. We shall show that there exist left stochastic matrices $X'$ and $Y'$ such that  for all $b \in B$, for all $R(\textbf{a}) \in \mathcal{C}_\bfa$, and for all $\ell \in L$ there exists $\hat{\ell}=\hat{\ell}(b,R(\ba),\ell) \in L_\ell$ such that $XA^\ell[b,R(\ba)] \leq B^{\hat{\ell}}Y[b,R(\textbf{a})]$. Assuming that this holds, again it follows by the induction hypothesis that $XX'$, $YY'$ are left stochastic matrices such that $XX'M_{\bfx_0}^\ell \leq \sum_{\ell' \in L_\ell} M_{\bfx_n}^{\ell'} Y Y' $ for all $ \ell \in L$.

Let $h$ be a homomorphism from $\bfa$ to $\bfb$. We define $X'[b,a]=1$ if $b=h(a)$ and $X'[b,a]=0$ otherwise. Similarly, we set $Y'[R^B(\bb),R^A(\ba)]=1$ if $\bb=h(\ba)$ and $R^B=R^A$ and $Y'[R^B(\bb),R^A(\ba)]=0$ otherwise. It is easy to see that $X'$ and $Y'$ are left stochastic. Now let $\ell =(S,R) \in L$, $b \in B$ and $R(\textbf{a}) \in \mathcal{C}_\bfa$. If $M^\ell_\bfa[a,R(\ba)]=0$ for all $a \in A$ then $XM^\ell_\bfa[b,R(\ba)]=0$ and there is nothing to prove. So we can assume that there is $a \in A$ such that for all $i \in [\arity(R)]$, $\ba[i] =a$ if and only if $i \in S$. Then we have that $XM^\ell_\bfa[b,R(\ba)]=1$ if $b=h(a)$, and $XM^\ell_\bfa[b,R(\ba)]=0$ otherwise. Again in the latter case there is nothing to prove so let us assume that $b=h(a)$. It follows that $h(\ba)[i]=b$ for all $i \in S$ and hence there exists $\hat{\ell}=(R,S')$ with $S \subseteq S'$ such that $M^{\hat{\ell}}_\bfb[b,R(h(\ba))]=1$, which completes the proof.

It only remains to prove $(\ref{1SA1Item}) \implies (\ref{4SA1Item})$. 
Assume that $\bfa$ and $\bfb$ satisfy $(\ref{1SA1Item})$. Further, we can assume that there exists an integer $m>0$ such that all variables in the feasible solution of $\sa^1(\bA,\bB)$ take rational values of the form $n/m$ for some integer $n$. Let $Y$ be the set of all tuples $((b_1,c_1),\dots,(b_m,c_m))\in (B\times[m])^m$ satisfying the following conditions:
\begin{itemize}
\item $(b_i,c_i)\neq (b_{i'},c_{i'})$ for every $i\neq i'\in [m]$ (i.e, the tuple has no repeated elements), and
\item for every $i\in [m]$, $c_i$ is at most $|\{j\in [m] \mid b_j=b_i\}|$.
\end{itemize}
In other words, $Y$ is the set of tuples that can be obtained if in every tuple $(b_1,\dots,b_m)\in B^m$ we replace all occurrences, say $n$, of any symbol $b\in B$ by $n$ `copies' $(b,1),\ldots,(b,n)$, in any possible order.
Let $\Pi:Y\rightarrow B^m$ be the function mapping $((b_1,c_1),\dots,(b_m,c_m))$ to $(b_1,\dots,b_m)$.

Let $X=A\times Y$. For every permutation $\tau$ on $[m]$ and any $m$-ary tuple $\bz=(z_1,\dots,z_m)\in Z^m$ where $Z$ is any arbitrary set we shall use $\bz\circ \tau$ to denote the tuple $(z_{\tau(1)},\dots,z_{\tau(m)})$. This notation is justified by the fact that we can see formally $\bz$ as a mapping from $[m]$ to $Z$. For any $x =(a, \textbf{y}) \in X$, we shall abuse notation and write $x \circ \tau$ to denote $(a, \by \circ \tau)$. 

For any two $\textbf{z},\textbf{z}'$ in any of the sets $B^m$, $Y$ or $X$, we shall write $\bz \sim \bz'$ iff there exists some permutation $\tau$ on $[m]$ such that $\bz'=\textbf{z} \circ \tau$.

For every multiset $U=\{\{ b_1,\dots,b_m\}\}$ of $m$ elements from $B$, we shall fix one arbitrary element of $Y$, denoted $\langle U \rangle$, satisfying that the multiset of variables in $\Pi(\langle U \rangle)$ coincides with $U$.
We shall abuse slightly notation and for every $\bb=(b_1,\dots,b_m)\in B^m$
we shall also use $\langle \bb\rangle$ to denote $\langle \{\{ b_1,\dots,b_m\}\}\rangle$. Clearly if $\bb \sim \bb'$ then $\langle \bb \rangle = \langle \bb' \rangle$. Moreover, for every $\bb\in B^m$ we shell denote by $\by^\bb$ the tuple $((b_1,c_1),\ldots,(b_m,c_m))\in Y$ obtained by choosing $c_i < c_j$ whenever $b_i=b_j$ and  $i < j$. Notice that $\Pi(\by)=\bb$ and $\by^\bb \sim \langle \bb\rangle$.

Now we shall show how to construct two
structures $\xone$ and $\xtwo$ which satisfy condition (\ref{4SA1Item}). The domain of both $\xone$ and $\xtwo$ is $X$. 
The constraints of $\xone$ and $\xtwo$ are constructed as follows. Let $J$ be the set 
of all triplets $(R,T,\ba)$ where $R\in\sigma$, $\ba\in A^{\arity(R)}$, and $T$ is a multiset of size $m$ of tuples in $R(\bB)$ such that for every $s,s'\in[\arity(R)]$:
\[\ba[s]=\ba[s']\Rightarrow \forall \bt\in T (\bt[s]=\bt[s']).\]
In other words, $T$ only contains tuples $\textbf{t} \in R(\bfb)$ such that $\textbf{t}=f(\textbf{a})$ for some function $f:\{\textbf{a}\} \to B$. 
Then for every $(R,T,\ba)\in J$ and $d\in[2]$, we define $Q^d_j$ to be the set of $m!$ constraints obtained in
the following way. Fix any arbitrary ordering 
$\bt_1,\dots,\bt_m$ of the tuples in $T$ and let $\bb_1\dots,\bb_r\in B^m$, $r=\arity(R)$ such
that $\bb_s[i]=\bt_i[s]$ for every $i\in[m]$ and $s\in[r]$. 
Then, for every permutation $\tau$ on $[m]$, we include constraint
$R((a_1,\langle \bb_1\rangle)\circ \tau,\dots,(a_r,\langle \bb_r\rangle)\circ\tau)$ 
in $Q_j^1$
and 
constraint 
$R((a_1,\by^{\bb_1})\circ\tau,\dots,(a_r,\by^{\bb_r})\circ\tau)$ in $Q_j^2$, where $\textbf{a}=(a_1,...,a_r)$.

Finally, we define $C_{\bX_d}$, $d\in[2]$ to be $\cup_{j\in J} Q^d_j$ (note that for each $d \in [2]$, all sets $Q^d_j$, $j\in J$ are disjoint). To complete the proof it only remains to show that $\xone$ and $\xtwo$ satisfy condition $(\ref{4SA1Item})$.

\begin{claim} \label{cla:2}
$\bfa$ is homomorphic to $\xone$. 
\end{claim}
\begin{claimproof}{1}
Let $p$ be a feasible solution of $\saone$. We define a mapping $h: A \to X$  by setting $h(a)=(a,\langle \textbf{c}_a\rangle)$, where $\textbf{c}_a\in B^m$ is any tuple satisfying that every element $b \in B$ occurs exactly $m \cdot p_a(b)$ times. We shall show that $h$ is a homomorphism from $\bfa$ to $\xone$. 
Let $R\in \sigma$, $\textbf{a}=(a_1,\ldots,a_r) \in \RA$ where $r=\arity(R)$, and consider the multiset $T=\{\textbf{t}_1,\ldots,\textbf{t}_m\}$ of tuples in $\RB$ obtained by picking each tuple $\textbf{t}_i=f_i(\textbf{a}) \in \RB$ exactly $m \cdot p_{R(\textbf{a})}(f_i)$ times (note that we are using implicitly (\ref{eq:SA4}) to guarantee the existence of $T$). For every $s,s'\in [r]$ satisfying $a_s=a_{s'}$ it follows from the construction of $T$ that $\bt[s]=\bt[s']$ for every $\bt\in T$, and hence $(R,T,\ba)\in J$. Let $\bt_1,\dots,\bt_m$ be the
ordering of the elements in $T$ associated to $(R,T,\ba)$ in the construction of $\xone$ and $\bb_1,\dots,\bb_r$ be the tuples obtained by setting $\textbf{b}_s[i]=\textbf{t}_i[s]$ for $i \in [m]$ and $s \in [r]$. Then, it follows that $\xone$ contains
the constraint $R((a_1,\langle \bb_1\rangle),\dots,(a_r,\langle \bb_r\rangle))$. It follows again from (\ref{eq:SA3})
that for every $s\in [r]$ $\textbf{c}_{a_s}\sim \bb_s$ and hence $(a_s,\langle \bb_s\rangle)$ is precisely $h(a_s)$ as desired.
\end{claimproof}

\begin{claim} 
$\xtwo$ is homomorphic to $\bB$. 
\end{claim}
\begin{claimproof}{2}
Let $h:X\rightarrow B$ be the function mapping any $(a,\by)$ to $\pr_{1}(\Pi(\by))$. It is immediate that $h$ defines
a homomorphism from $\xtwo$ to $\bB$.
\end{claimproof}

\begin{claim} \label{cla:3}
$\xone$ and $\xtwo$ have a common equitable partition.
\end{claim}

\begin{claimproof}{3}
We shall prove that $(P,Q^1)$ and $(P,Q^2)$ define a common equitable partition of $\xone$ and $\xtwo$ where $P=\{P_i \mid i\in I\}$
is the partition given by the equivalence relation $\sim$ on $X$ and $Q^d=\{Q^d_j \mid j \in J\}$, $d\in[2]$ are as defined in the construction of $\xone$ and $\xtwo$.

This follows immediately from the following fact. Let $d\in[2]$, $i\in I$, $j\in J$, and $\ell\in L$. Then exactly one of the following conditions holds:
\begin{enumerate}
\item There is no $x\in P_i$ and $C\in Q^d_j$ such that $M_{\bX_d}[x,C]=\ell$; \label{partition1item}
\item There exists a one-to-one correspondence between the elements of $P_i$ and $Q^d_j$
such that for every pair $(x,C)$ of associated elements, $M_{\bX_d}[x,C]=\ell$. \label{partition2item}
\end{enumerate} 
Furthermore, for every $i\in I$, $j\in J$, and $\ell\in L$, condition (\ref{partition1item}) holds for $d=1$ if and only if it holds for $d=2$. 

Let us prove it. Let $j=(R,T,\ba)$ and let $\bb_1,\dots,\bb_r$ and $\by^{\bb_1},\dots,\by^{\bb_r}$ be as in the construction of $Q_j^d$ (where $r=\arity(R))$.  We first
observe that for every $x\in P_i$, every $C=R(x_1,\dots,x_r)\in Q^d_j$, and every permutation $\tau$ on $[m]$,
$M_{\bX_d}[x,C]=\ell\Leftrightarrow M_{\bX_d}[x\circ \tau,C\circ \tau]=\ell$ where we use $C\circ \tau$ to denote $R(x_1\circ \tau,\dots,x_r\circ \tau)$.
It then follows that if (\ref{partition1item}) fails then (\ref{partition2item}) must hold.

Now it only remains to see that (\ref{partition1item}) holds for $d=1$ if and only if (\ref{partition1item}) holds for $d=2$.
Clearly, this follows immediately if the relation symbol in $\ell$ is different from $R$, so we can
assume that $\ell=(R,S)$ for some $S\subseteq [r]$. We then note that if there is some pair $(x,C)$ violating (1), then $M_{\bX_d}[x\circ \tau,C\circ \tau] = \ell$ for all permutations $\tau$ on $[m]$, and hence $C$ can be chosen to be any constraint in $Q^d_j$. Hence, if we choose $R((a_1,\langle \bb_1\rangle),\dots,(a_r,\langle \bb_r\rangle))$ for $d=1$ and $R((a_1,\by^{\bb_1}),\dots,(a_r,\by^{\bb_r}))$ for $d=2$, in order to complete
the proof it is enough to show that for every $s,s'\in[r]$, $(a_{s},\langle \bb_{s}\rangle)=(a_{s'},\langle \bb_{s'}\rangle)\Leftrightarrow (a_s,\by^{\bb_s})=(a_{s'},\by^{\bb_{s'}})$. 

The direction ($\Leftarrow$) is immediate. For the direction $(\Rightarrow)$ assume that $s,s'$ satisfy the left-hand side. Since $a_s=a_{s'}$ it follows that $\bb_s=\bb_{s'}$. Since $\by^{\bb_s}$ and $\by^{\bb_{s'}}$ are determined in a unique way from $\bb_s$ and $\bb_{s'}$
respectively, we are done.\end{claimproof}

%% file: 3Applicatons-arxiv.tex
\section{Some Applications} \label{sec:applications}

As an immediate consequence of Theorem \ref{thm:SAk} we obtain that feasibility of the $k^{th}$ Sherali-Adams relaxation of the homomorphism problem is closed under $\eqwl[k]$. 
 
 \begin{corollary}
Let $\bA$, $\bA'$ and $\bB$ be $\sigma$-structures and suppose that $\bA\eqwl[k]\bB$. Then, $\sak$ is feasible  if and only if $\sa^k(\bA',\bB)$ is feasible.
\end{corollary}

We note that if the maximum arity on the signature $\sigma$ is at most $k$, then the previous corollary can be alternatively stated in the following way: if $\bA$ and $\bA'$ are $C^k$-equivalent then $\sak$ is feasible if and only if $\sa^k(\bA',\bB)$ is feasible. Then, one could be tempted to use this observation to transfer results from LP relaxations to logical definability. In particular one could infer that if $\text{SA}^k$ decides correctly
$\csp(\bB)$ then it is definable in the logic $C^k_{\infty,\omega}$ which is the extension of $C^k$ consisting of all formulas made from atomic formulas and equality by means of finitary and infinitary conjunctions, negations, and standard and counting quantifiers. However, this is of limited interest as it follows by combining \cite{AtseriasBD09} and \cite{Barto2016Collapse} that $\csp(\bB)$
would also be definable in the much weaker logic $L^{\max(k,3)}_{\infty,\omega}$ where counting quantifiers are not allowed. Consequently, the previous corollary is most likely to find applications in obtaining lower bounds on the Sherali-Adams rank for concrete instances of CSP.

However, the principal novelty in our result is precisely the opposite direction, which leads to an alternative combinatorial characterization of the Sherali-Adams relaxation. A concrete application is the answer to the following question: for which structures $\bB$ is $\csp(\bB)$ closed under $\eqwl[1]$-equivalence? This question arises in the context of the distributed CSP \cite{Butti2021} where the variables and constraints of an instance are distributed among agents which communicate with each other by sending messages through fixed communication channels. In fact, the connection between the Weisfeiler-Leman algorithm and distributed computation goes back to the influential paper of Angluin on networks of processors \cite{Angluin1980Local}. For the distributed CSP, one of the most natural configurations for the communication network is essentially identical to the factor graph \cite{Fioretto2018}. It then follows, under some technical requirements (agent anonymity and synchronicity) that any distributed message passing algorithm will necessarily behave in an identical manner on every two input instances that are $\eqwl[1]$-equivalent and,  hence, it follows that $\csp(\bB)$ can only be solved by a distributed algorithm if $\csp(\bB)$ is closed under $\eqwl[1]$-equivalence.

This question was already answered in \cite{Butti2021} where it was shown that $\csp(\bB)$ is closed under $\eqwl[1]$ if and only if $\bB$ has symmetric polymorphisms of all arities, where a $k$-ary symmetric polymorphism of a $\sigma$-structure $\bB$ is any homomorphism $h$ from $\bB^k$ to $\bB$ that is invariant under the permutation of its arguments (the reader can safely ignore the definition of symmetric polymorphism as we will be using it as a black-box). The proof in \cite{Butti2021} makes use of a result from \cite{Kun2012} stating that $\csp(\bB)$ has symmetric polymorphisms of all arities if and only if it is solvable by an LP relaxation known as the basic linear programming relaxation (BLP). Although BLP is slightly different from $\sa^1$, both coincide over instances $(\bA,\bB)$ where $\bA$ has no loops (i.e, every constraint has no repeated elements). It is then very easy to obtain
the following characterization:

\begin{lemma}
Let $\bB$ be a fixed finite $\sigma$-structure. The following are equivalent:
\begin{enumerate}
    \item $\csp(\bB)$ is closed under $\eqwl[1]$-equivalence; \label{1application}
    \item $\sa^1$ decides $\csp(\bB)$; \label{2application}
    \item $\blp$ decides $\csp(\bB)$; \label{3application}
    \item $\bB$ has symmetric polymorphisms of all arities. \label{4application}
\end{enumerate}
\end{lemma}

\begin{proof}[Proof (Sketch)] $(\ref{1application})\Leftrightarrow (\ref{2application})$ follows from Theorem \ref{thm:SA1} and $(\ref{3application})\Leftrightarrow (\ref{4application})$ from \cite{Kun2012}. Hence, it is only necessary to verify $(\ref{2application})\Leftrightarrow (\ref{3application})$.  We use the following fact which follows from the Sparse Incomparability Lemma \cite{Nesestril1989}: for every instance $\bA$ of $\csp(\bB)$, there exists a structure $\bA'$ with no loops such that $\bA'\rightarrow \bA$ and $\bA\rightarrow \bB$ iff $\bA'\rightarrow \bB$. Now, assume that $\blp$ does not solve $\csp(\bB)$. This means that there exists a structure $\bA$ not homomorphic to $\bB$ and such that  $\blp(\bA,\bB)$ is feasible. Now, let $\bA'$ be the structure given by the Sparse Incomparability Lemma. Since $\bA'\rightarrow \bA$ it follows that $\blp(\bA',\bB)$ is feasible, and, since $\bA'$ has no loops, $\sa^1(\bA',\bB)$ is feasible as well. Since $\bA'$ is not homomorphic to $\bB$ it follows that $\sa^1$ does not solve $\csp(\bB)$. The same argument can be used to show the converse although it is not necessary as it also follows immediately by comparing the inequalities of $\sa^1$ and $\blp$.
\end{proof}

%% file: Appendix-arxiv.tex
\appendix 

\section{Applying the SA method to the Homomorphism problem}
\label{appendix:SA}

Here we shall show that the family of relaxations $\sa^k$ considered in the present paper is 
closely interleaved with the system of relaxations obtained by applying the SA method to a natural choice of initial polytope $P$.

Let $\bA$ and $\bB$ be $\sigma$-structures. We define polytope $P=P(\bA,\bB)$ using a system of inequalities. The variables of the system are $x_{a,b}$ for each $a\in A$ and $b\in B$. Each variable must take a value in the range $[0,1]$. We remark that by fixing some arbitrary ordering on the variables in $x_{a,b}$ we can represent any assignment on the variables $x_{a,b}$ with a tuple $\bx\in\mathbb{R}^n$ with $n=|A|\cdot |B|$. Therefore we shall abuse notation and use $\bx_{a,b}$ to refer to the value in $\bx$ corresponding to variable $x_{a,b}$.

The variables are subject to the following inequalities. 
\begin{equation}\label{eq:blp2}
\sum_{b\in B} x_{a,b}=1 \text{ for every $a\in A$},
\end{equation}

\begin{equation}\label{eq:blp3}
\sum_{a\in\{\ba\}} x_{a,f(a)}\leq |\{\ba\}|-1
\begin{array}{l}
\text{ for each $R\in\sigma$, $\ba\in \RA$, }\\
\text{and $f:\{\ba\}\rightarrow B$ with $f(\ba)\not\in \RB$.}
\end{array}
\end{equation}

Note that if $h$ is a homomorphism from $\bA$ to $\bB$ then the assignment setting $x_{a,h(a)}=1$ for every $a\in A$ and the rest of variables to zero is feasible.

Now let $P^k, k\geq 1$ be the sequence of polytopes obtained using the SA method. The next lemma shows that the sequence of relaxations defined by $\sa^k$ and $P^k$ are interleaved.

\begin{lemma} \label{le:SAkrPk}
Let $k\geq 1$ and let $r$ be the maximum arity of a relation in $\sigma$. Then
\begin{enumerate}
\item If $P^k\neq\emptyset$ and $r\leq k$ then $\sa^{k}$ is feasible. \label{1SAkrPk}
\item If $\sa^{k+r-1}$ is feasible then $P^k\neq\emptyset$  \label{2SAkrPk}
\end{enumerate}
\end{lemma}
\begin{proof}

(\ref{1SAkrPk}). Assume that $P^k\neq\emptyset$ and let $\by$ be a feasible solution of $P^k_L$. We shall construct a feasible solution of $\sa^k$. First, set every variable of the form $p_V(f)$ to $y_K$ where $K=\{(a,f(a)) \mid a\in V\}$. We first observe that this assignment satisfies (\ref{eq:SA1}) and (\ref{eq:SA2}). Indeed, let $U\subseteq A$ with $|U|<k$, let $f:U\to B$, and let $I=\{(u,f(u)) : u\in U\}$. Then, multiplying the equality (\ref{eq:blp2}) with $a\in A\setminus U$ by $\Pi_{i\in I} x_i$ and linearizing we obtain equality (\ref{eq:SA2}) for $U$, $f$,  and $V=U\cup\{a\}$. In this way we can obtain all equalities in (\ref{eq:SA2}) for $|U|+1=|V|$. We note here that the rest of equalities in (\ref{eq:SA2}) along all equalities in (\ref{eq:SA1}) can be obtained as a linear combination.

Secondly, let us set the rest of variables. For every $(\ba,R)\in \mathcal{C}_{\bA}$ and $f:\{\ba\}\to B$, set $p_{(\ba,R)}(f)$ to be $y_K$ where $K=\{(a,f(a)) \mid a\in\{\ba\}\}$ (note that we are using implicitly the fact that $r\leq k$). Then, (\ref{eq:SA3}) follows directly from (\ref{eq:SA2}). Finally, it only remains to show that (\ref{eq:SA4}) is also satisfied. Indeed, for every $f(\ba)\not\in \RB$ we obtain
equality $p_{(\textbf{a},R)}(f) = 0$ if we multiply (\ref{eq:blp3}) by the term $\Pi_{i\in K}$ and linearize.
We want to note that, in fact, (\ref{1SAkrPk}) also holds under the weaker assumption $r\leq k+1$, but the proof is slightly more involved.

(\ref{2SAkrPk}). Assume that $\sa^{k+r-1}$ is feasible. We construct a feasible solution $\by$ of $P^k_L$ as follows. For every $K \subseteq A \times B$ which satisfies $K=\{(a,f(a)) \mid a\in U\}$ for some $U\subseteq A$ with $|U|\leq k$ and $f:U \to B$, we set $y_K:=p_U(f)$. Otherwise, we set $y_K$ to zero. 

 Let us show that this assignment satisfies all inequalities in $P^k_L$. Let 
 \begin{equation}\label{eq:cyd}
 \bc^T \by\leq \bd
 \end{equation}
 be any inequality defining $P^k_L$. Since (\ref{eq:cyd}) is obtained by multiplying an inequality which contains at most $r$ variables by a term of at most $k-1$ variables, there 
exists a set $V\subseteq A$ with $|V|\leq r+k-1$ such that for every variable $y_K$ appearing in (\ref{eq:cyd}), $V$ satisfies
$K\subseteq V\times B$. Note that, by (\ref{eq:SA1}),  variables $p_V(g)$, $g:V \to B$ define a probability distribution. For every $g:V \to B$ in the support of this distribution, consider the assignment $\bx^g$ that sets $\bx^g_{v,b}=1$ if $v\in V$ and $b=g(v)$ and $x_{v,b}=0$ otherwise. 

Inequality (\ref{eq:cyd}) has been obtained by multiplying an inequality from (\ref{eq:blp2}) or (\ref{eq:blp3}) by a term and linearizing. We claim that in both cases, the inequality that has generated (\ref{eq:cyd}) is satisfied by $\bx^g$. If the inequality generating (\ref{eq:cyd}) is $\sum_{b\in B} x_{a,b}=1$ for some $a\in A$ this follows simply from the fact that $a\in V$. Assume now that (\ref{eq:cyd}) has been generated by inequality $\sum_{a\in\{\ba\}} x_{a,f(a)}\leq |\{\ba\}|-1$. In this case
note that $\{\ba\}\subseteq V$ and then the claim follows from  (\ref{eq:SA3}) and (\ref{eq:SA4}). This finalizes the proof of the claim.

Consequently, since $\bx^g$ is integral it follows that the assignment $\by^g$ defined as $\by^g_K=\Pi_{i\in K} \bx^g_i$ satisfies (\ref{eq:cyd}). Finally, note that if we set $\alpha^g=p_V(g)$, then for every $K\subseteq V\times B$, $\by_K$ is precisely
given by the convex combination $\sum_g \alpha^g \by^g_K$.\end{proof}

\section{Proof of Lemma \ref{le:starkSherali}}

\lestarkSherali*

\begin{proof}[Proof (Sketch)]
The proof is purely syntactical although it is convenient to massage first a bit the LP formulations $\sak$ and $\sastar$. We shall refer to the solutions of $\sak$ and $\sastar$ by appropriately indexed sets of variables $p$, $q$ respectively. 

\begin{itemize}
\item In $\sak$, it follows from (\ref{eq:SA4}) that we can safely replace all variables $p_{R(\ba)}(f)$ with $f(\textbf{a}) \not\in R(\bB)$ by $0$.
\item In $\sastar$ we are required a bit more of work. First, for each $j\leq k$ and each $\ba\in A^j$, it follows from conditions (\ref{eq:SA3}) and (\ref{eq:SA4}) for
$T_{j,S}$ ($S\subseteq[j])$ that for every $x$ in $\bstark$, $q_{\ba}(x)$ must take value $0$ unless $x=f(\ba)$ for some function $f:\{\ba\}\rightarrow B$. Hence, in a first stage
we set $q_{\ba}(x)$ to zero for each $j\leq k$, each $\ba\in A^j$
and each $x$ that is not a tuple of the form $f(\ba)$ for some function $f:\{\ba\}\rightarrow B$.

Furthermore, it follows from condition (\ref{eq:SA3}) for $\tjii$ that $q_{\ba}(f(\ba))=q_{\ba'}(f(\ba'))$ for every $\ba$, $\ba'$ satisfying $\{\ba\}=\{\ba'\}$ and every $f:\{\ba\}\rightarrow B$. Hence, in a second stage,
for each $V\subseteq A$ with $|V|\leq k$ and every $f:V\rightarrow B$ we identify all variables $q_{\ba}(f(\ba))$ which satisfy 
$\{\ba\}=V$. 

Then, consider now the variables of the form $q_{R(\ba)}(x)$, $x\in\bstark$. It
follows from conditions (\ref{eq:SA3}) and (\ref{eq:SA4}) for $R_S$ ($S\subseteq[\arity(R)])$ that
$q_{R(\ba)}(x)$ must be set to $0$ unless $x=R(f(\ba))$ for some function $f:\{\ba\}\rightarrow B$.

The other variables in $\sastar$ are of the form $q_C(f)$ where $C\in \mathcal{C}_{\astarkbf}$. As we shall see they can always safely be identified with some of the other variables. Let us start first with the case in which $C$ is a unary constraint. If $C=T_{j,S}(\ba)$ or $C=R_S(\ba)$, then it follows from (\ref{eq:SA3}) that $q_C(f)=q_{\ba}(f(\ba))$. Assume
now that $C$ is a binary constraint, that is $C=\tjii(\ba,\pii \ba)$ or $C=R_{\bf i}(\ba,\pii \ba)$. It follows again from 
(\ref{eq:SA3}) that $q_C(f)=q_{\ba}(f(\ba))$ 
\end{itemize}

Now we are ready to prove the lemma. In particular, consider the following one-to-one correspondence between
the assignments in $\sak$ and $\sastar$:
\begin{itemize}
\item Every variable $p_V(f)$ in $\sak$ is assigned as variable $q_{\ba}(f(\ba))$ in $\sastar$ where $\ba$ is any tuple satisfying $\{\ba\}=V$.
\item Every variable $p_{R(\ba)}(f)$ in $\sak$ is assigned as variable $q_{R(\ba)}(R(f(\ba)))$ in $\sastar$.
\end{itemize}

It is not difficult to see that this correspondence preserves feasibility.\end{proof}

\section{Proof of Theorem \ref{le:digraph}}

\ledigraph*

Before starting we note that two structures $\bA$ and $\bB$ have a common equitable partition if
there is an equitable partition $(\{P_i \mid i\in I\},\{Q_j \mid j\in J\})$ of the disjoint union
$\bA\cup \bB$ such that 
\begin{enumerate}
    \item $|P_i\cap A|=|P_i\cap B|$ for every $i\in I$, and \label{1conditionPart}
    \item $|Q_j\cap \mathcal{C}_{\bA}|=|Q_j\cap \mathcal{C}_{\bB}|$ for every $j\in J$. \label{2conditionPart}
\end{enumerate}
Note that since $(P,Q)$ is equitable, then any of conditions (\ref{1conditionPart}) or (\ref{2conditionPart}) implies the other. We might freely switch between the two alternative characterizations of common equitable partition. 

We also need a few additional definitions before embarking on the proof. 
We denote by $J_V$  the $V \times V$ matrix whose entries are all ones and we use $\oplus$ to denote direct sums of matrices. Let $M$ be a $V\times W$ matrix and consider two subsets $S_{V} \subseteq V$, $S_{W} \subseteq W$. The restriction of $M$ to $S_{V}$ and $S_{W}$, denoted by $M[S_{V},S_{W}]$, is the matrix obtained by removing from $M$ all the rows that do not belong in $S_{V}$ and all the columns that do not belong in $S_{W}$.

A matrix $M \in \mathbb{R}^{V\times W}$ is \textit{decomposable} if there exist a partition $V_1,V_2$ of $V$ 
and a partition $W_1,W_2$ of $W$ such that for every different $i,j\in\{1,2\}$ and every $v\in V_i$ and $w\in W_j$, $M_{v,w}=0$. In this case we can write $M=M_1\oplus M_2$ where $M_i=M[V_i,W_i]$. Otherwise, $M$ is said to be \textit{indecomposable}.

Let $X\in[0,1]^{V\times V}$ be a doubly stochastic matrix. Note that $X$ has a unique decomposition $X=\oplus_{i\in I} X_i$ where each $X_i$ is an indecomposable doubly stochastic matrix. The \textit{row partition} of $X$ is defined to be the partition of $V$ into classes $P_i,i\in I$ where $P_i$ contains the rows of $X_i$, and the \textit{column partition} is defined in an analogous manner.

We also need the following lemma (see for example \cite{Scheinerman2011fractional}).

\begin{lemma} \label{le:indecomposable}
Let $X\in \mathbb{R}^{V \times V}$, $Y\in \mathbb{R}^{W \times W}$ be doubly stochastic indecomposable matrices.
\begin{enumerate}
    \item Let  ${\bf a},{\bf b}\in\mathbb{R}^{V}$ such that $X\cdot {\bf a}={\bf b}$ and $X^T\cdot {\bf b}={\bf a}$. Then, there exists $c\in\mathbb{R}$ such that \[{\bf a}={\bf b}=c\cdot {\bf 1.}\] \label{1indecomposableItem}
    \item Let $M_1,M_2\in\mathbb{R}^{V\times W}$ such that $XM_1=M_2Y$ and $M_1Y^T=X^T M_2$. \label{miitem}
Then, there exists $c,d\in\mathbb{R}$ such that the following identities hold:
\[M_1{\bf 1}=M_2{\bf 1}=c\cdot{\bf 1} \hspace{2cm} {\bf 1}^T M_1={\bf 1}^T M_2=d \cdot {\bf 1}^T\] \label{2indecomposableItem}
\end{enumerate}
\end{lemma}

\begin{proof}
Item (\ref{1indecomposableItem}) is Theorem 6.2.4 (ii) from \cite{Scheinerman2011fractional}. We enclose a proof of item (\ref{2indecomposableItem}). Let ${\bf a}=M_1\cdot {\bf 1}$ and ${\bf b}=M_2\cdot {\bf 1}$. We have $XM_1\cdot {\bf 1}=M_2 Y\cdot{\bf 1}=M_2\cdot {\bf 1}$ since $Y$ is doubly stochastic, which implies $X\cdot {\bf a}={\bf b}$. Similarly, we have $M_1\cdot {\bf 1}=M_1 Y^T\cdot {\bf 1}=X^T M_2\cdot {\bf 1}$ which implies ${\bf a}=X^T\cdot {\bf b}$. Then, we apply item (\ref{1indecomposableItem}) to deduce that there exists $c \in \mathbb{R}$ such that $M_1{\bf 1}=M_2{\bf 1}=c\cdot{\bf 1}$. The other identity is proven in an analogous way.
\end{proof}

 $(\ref{1digraphItem})\Rightarrow (\ref{3digraphItem})$. Let $X$ and $Y$ be doubly stochastic matrices satisfying (\ref{1digraphItem}). Let $X=\oplus_{i\in I} X_i$, $Y=\oplus_{j\in J} Y_j$ be decompositions of $X$ and $Y$. Denote by $(P^{\bfa},P^{\bfb})$ the column and row partitions of $X$ respectively and by $(Q^{\bfa},Q^{\bfb})$ the column and row partitions of $Y$.
Then, the restrictions of $M_{\bfa}^{\ell}$ and $M_{\bfb}^{\ell}$ to $(P^{\bfa}_{i},Q^{\bfa}_{j})$, $(P^{\bfb}_{i},Q^{\bfb}_{j})$ respectively  satisfy 
\[X_i M_{\bfa}^{\ell}[P^{\bfa}_{i},Q^{\bfa}_{j}]= M_{\bfb}^{\ell}[P^{\bfb}_{i},Q^{\bfb}_{j}] Y_j,\]
\[M_{\bfa}^{\ell}[P^{\bfa}_{i},Q^{\bfa}_{j}]Y_j^T= X_i^T M_{\bfb}^{\ell}[P^{\bfb}_{i},Q^{\bfb}_{j}].\]

Hence it follows from Lemma \ref{le:indecomposable} that there exist $c^{\ell}_{i,j}$, $d^{\ell}_{i,j}$ such 
that 
\begin{align*}
&M_{\bfa}^{\ell}[P^{\bfa}_{i},Q^{\bfa}_{j}]\cdot{\bf 1}=M_{\bfb}^{\ell}[P^{\bfb}_{i},Q^{\bfb}_{j}]\cdot{\bf 1}=c^{\ell}_{i,j}\cdot {\bf 1} \\
&{\bf 1}^T \cdot M_{\bfa}^{\ell}[P^{\bfa}_{i},Q^{\bfa}_{j}]={\bf 1}^T \cdot M_{\bfb}^{\ell}[P^{\bfb}_{i},Q^{\bfb}_{j}]={\bf 1}^T\cdot d^{\ell}_{i,j} 
\end{align*}
which is equivalent to conditions (\ref{eq:eqpartQ}) and (\ref{eq:eqpartP}) showing that partitions $(P^{\bfa},Q^{\bfa})$ and $(P^{\bfb},Q^{\bfb})$ have the same parameters.

$(\ref{3digraphItem})\Rightarrow (\ref{2digraphItem})$. Assume that $(\{P^{\bfa}_i \mid i\in I\},\{Q^{\bfa}_j \mid j\in J\})$ and $(\{P^{\bfb}_i \mid i\in I\},\{Q^{\bfb}_j \mid j\in J\})$ define a common equitable partition of $\bfa$ and $\bfb$. We shall prove that any two elements that are in the same set of the partition of $\bfa$ and $\bfb$ must have the same iterated degree. That is, we show by induction on $k$ that for all $k\geq 0$, $\delta_k^{\bfa}(a)=\delta_{k}^{\bfb}(b)$ whenever there exists $i\in I$ such that $a\in P^{\bfa}_i$ and $b\in P^{\bfb}_i$ (the case $R(\textbf{a}) \in Q^{\bfa}_j,R(\textbf{b}) \in Q^{\bfb}_j$ is analogous).
The base case ($k=0$) is immediate. For the inductive case, assume that the statement holds for $k-1$. 
Let $(\ell,\delta)$ be any arbitrary element in $\delta^{\bfa}_k(a)$. We shall show that it has the same multiplicity in $\delta^{\bfa}_k(a)$ and in $\delta^{\bfb}_{k}(b)$. By the inductive hypothesis it follows that there exists $J_{\delta}\subseteq J$ such that 
\[\{R(\textbf{a}) \in \mathcal{C}_\bfa \mid \delta^{\bfa}_{k-1}(R(\textbf{a}))=\delta\}=\bigcup_{j\in J_{\delta}} Q_{j}^{\bfa}\]
\[\{R(\textbf{b})\in  \mathcal{C}_\bfb \mid \delta^{\bfb}_{k-1}(R(\textbf{a}))=\delta\}=\bigcup_{j\in J_{\delta}} Q_{j}^{\bfb}\]
Then, the multiplicity of $(\ell,\delta)$ in $\delta^{\bfa}_k(a)$ is
\begin{equation}\label{eq:multiplicity}
|\{R(\textbf{a}) \in\bigcup_{j\in J_{\delta}} Q_{j}^{\bfa} \mid  M_{\bfa}[a,R(\textbf{a})]=\ell\}|.
\end{equation}

Since $(P^{\bfa},Q^{\bfa})$ is an equitable partition it follows that (\ref{eq:multiplicity}) is equal to $\sum_{j\in J_{\delta}} c_{i,j}^{\bfa,\ell}$
where $c_{i,j}^{\bfa,\ell}$ are the parameters of the partition.

It can analogously be shown that the the multiplicity of $(\ell,\delta)$ in $\delta^{\bfb}_k(b)$ is
$\sum_{j\in J_{\delta}} c_{i,j}^{\bfb,\ell}$.  Since $(P^{\bfa},Q^{\bfa})$ and $(P^{\bfb},Q^{\bfb})$ define
a common equitable partition it follows that $c^{\bfa,\ell}_{i,j}=c^{\bfb,\ell}_{i,j}$ for every $i\in I,j\in J$ and we are done.

$(\ref{2digraphItem})\Rightarrow (\ref{3digraphItem})$. Assume that $\bfa$ and $\bfb$ have the same iterated degree sequence, let $\delta_k^\bfa$ and $\delta_k^\bfb$ be a fixed point. Let $P=\{P_i \mid i\in I\}$ and $Q=\{Q_j \mid j\in J\}$ be a partition of $A\cup B$ and $\mathcal{C}_{\bA}\cup \mathcal{C}_{\bB}$ respectively where each class contains precisely all elements with the same $k^{th}$ degree. It is easy to 
verify that $(P,Q)$ defines a common equitable partition.

$(\ref{3digraphItem})\Rightarrow (\ref{1digraphItem})$. Assume that 
$(\{P^{\bfa}_i \mid i\in I\},\{Q^{\bfa}_j \mid j\in J\})$ and $(\{P^{\bfb}_i \mid i\in I\},\{Q^{\bfb}_j \mid j\in J\})$ define a common equitable partition. For ease of notation it is convenient that matrices $M^{\ell}_\bfa$ and $M^{\ell}_\bfb$
are indexed by the same sets of rows $V$ and columns $W$, which can be done by fixing a one-to-one correspondence
between $A$, $B$ and $V$ and similarly between $\mathcal{C}_{\bA}$, $\mathcal{C}_{\bB}$, and $W$. Furthermore, the correspondence can be established
in such a way that under this correspondence $P^{\bfa}$ and $P^{\bfb}$ become the same partition $P$ over $V$, and $Q^{\bfa}$ and $Q^{\bfb}$ become the same partition $Q$ over $W$.

Let $M^{X}_{i} = \frac{1}{|P_{i}|} J_{P_i}$, $M^{Y}_{j} = \frac{1}{|Q_{j}|} J_{Q_j}$, and define $X = \oplus_{i \in I} M^{X}_{i}$ and $Y = \oplus_{j \in J} M^{Y}_{i}$. Clearly $X$ and $Y$ are doubly stochastic. It remains to show that $XM_{\bfa}^{\ell}=M_{\bfb}^{\ell}Y$ and $M_{\bfa}^{\ell}Y^{T}=X^{T}M_{\bfb}^{\ell}$ for all labels $\ell \in L$. 
Now since $(P,Q)$ is an equitable partition, it follows that for every $i \in I$, $j \in J$, and $\ell \in L$ there exist parameters $c^{\ell}_{ij}$ and $d^{\ell}_{ji}$ which satisfy conditions (\ref{eq:eqpartQ}) and (\ref{eq:eqpartP}).

Then for all $i \in I$, $j \in J$ and $\ell \in L$ it holds that 
\begin{equation*}
|P_{i}| c^{\ell}_{ij} = |Q_{j}|d^{\ell}_{ji},
\end{equation*}
and that the sum of the elements of the respective $P_{i}\times Q_{j}$ portions of $M_{\bfa}^{\ell}$ and $M_{\bfb}^{\ell}$ are equal. Now let $b \in P_{i} \cap B$ and $R(\textbf{a}) \in Q_{j} \cap \mathcal{C}_{\bfa}$. Then 
\begin{align*}
    (XM_{\bfa}^{\ell})[b,R(\textbf{a})] = \frac{1}{|P_{i}|} d_{ji}^{\ell} = \frac{1}{|Q_{j}|} c_{ij}^{\ell} = (M_{\bfb}^{\ell}Y)[b,R(\textbf{a})], 
\end{align*}
showing that $XM_{\bfa}^{\ell} = M_{\bfb}^{\ell}Y$ for all labels $\ell \in L$ as required, and similarly, noting that $X^{T}=X$ and $Y^{T}=Y$, we obtain that $M_{\bfa}^{\ell}Y^{T} = X^{T}M_{\bfb}^{\ell}$ for all $\ell \in L$ too.

$(\ref{1digraphItem})\Rightarrow (\ref{5digraphItem})$. Assume that $X$ and $Y$ satisfy (\ref{1digraphItem}). We shall prove that $X$ satisfies (\ref{5digraphItem}).

Let $R$ be the predicate corresponding to the edge relation. We define two matrices $U_{\bA}$, $Z_{\bA}$ where $U_{\bA}=(M_{\bA}^{(1,R)}+M_{\bA}^{(2,R)})/\sqrt{2}$ and $Z_{\bA}=2M_{\bA}^{(1,R)}/\sqrt{2}$.
Then we have that $\na=U_{\bA} U_{\bA}^T-Z_{\bA} Z_{\bA}^T$. We similarly obtain $\nb=U_{\bB} U_{\bB}^T- Z_{\bB}Z_{\bB}^T$ if we define $U_{\bB}$ and $Z_{\bB}$ accordingly. 
Note that the identities of (\ref{1digraphItem}) are still satisfied if we replace $M_{\bA}^{\ell}$ and $M_{\bB}^{\ell}$ by $U_{\bA}$
and $U_{\bB}$ or $Z_{\bA}$
and $Z_{\bB}$ respectively.

It then follows that
\[X U_{\bA} U_{\bA}^T= U_{\bB} Y U_{\bA}^T = U_{\bB} U_{\bB}^T X. \] and \[X Z_{\bA} Z_{\bA}^T= Z_{\bB} Y Z_{\bA}^T = Z_{\bB} Z_{\bB}^T X. \] 
Therefore, $X \na = \nb X$ as required.

$(\ref{5digraphItem})\Rightarrow (\ref{3digraphItem})$. Let $R$ be the predicate expressing the edge relation. Let $X$ be a doubly stochastic matrix satisfying $X\na= \nb X$. Since $\na$ and $\nb$ are symmetric it also holds that $\na X^T= X^T \nb$.  Let $X=\oplus_{i\in I} X_i$ be a decomposition of $X$ and denote by $\{P_i^{\bfa} \mid i\in I\}$ and $\{P_i^{\bfb} \mid i\in I\}$ the column and row partitions of $X$. Applying the same reasoning as in $(\ref{1digraphItem})\Rightarrow (\ref{3digraphItem})$ it follows that for every $i,i'\in I$, there exists $c_{i,i'}$ such that for every $\bD\in \{\bA,\bB\}$ and every $d\in P_i^{\bD}$
\[\{ d'\in P^{\bD}_{i'} \mid R(d,d')\in \mathcal{C}_{\bA}\}=c_{i,i'}  \]
It then easily follows that if we let $Q^{\bfa}$, $Q^{\bfb}$ be the partitions of $\mathcal{C}_\bfa$, $\mathcal{C}_\bfb$ respectively given by assigning two edges to the same partition class if the  partition classes of their vertices coincide then $(P^{\bfa},Q^{\bfa})$ and $(P^{\bfb},Q^{\bfb})$ define
a common equitable partition of $\bA$ and $\bB$.

The proof of $(\ref{3digraphItem}) \iff (\ref{4digraphItem})$ requires some more work. We shall use the following technical lemma.

\begin{claim}
\label{cl:independent}
Let $n\geq 1$, let $X$ be a family of $n$-ary vectors of non-negative real numbers
such that for every $\ba,\bb\in X$, if the new vector $\ba\cdot\bb$ obtained by multiplying $\ba$ and $\bb$ component-wise
    is different from ${\bf 0}$ then it also belongs to $X$. 
Let $P_X=\{ i\in[n] \mid \forall \ba\in X(\ba[i]\neq 0)\}$ and $D_X=\{(i,j) \mid \exists \ba\in X (\ba[i]\neq \ba[j])\}$. Then
\begin{enumerate}
    \item There exists $\ba\in X$ such that $\ba[i]=0$ for every $i\not\in P_X$ and $\ba[i]\neq \ba[j]$ for every 
    $(i,j)\in D_X\cap (P_X\times P_X)$. \label{1independent}
    \item Furthermore if $D_X$ contains all non reflexive pairs in $[n]\times [n]$ then $X$ contains $n$ linearly independent vectors. \label{2independent}
\end{enumerate}
\end{claim}
\begin{claimproof}{4}
$(\ref{1independent})$. We shall prove by every subset $Y$ of $X$ there exists an assignment satisfying (\ref{1independent}) replacing $P_X$ by $P_Y$ and 
$D_X\cap (P_X\times P_X)$ by $D_Y\cap (P_Y\times P_Y)$. The proof is by induction on $|Y|$. The base case when $Y$ only contains a single vector is trivial. For the inductive case, let $Y=Z\cup\{\bb\}$ and let
$\ba$ be the tuple satisfying the claim for $Z$. It follows immediately that for $d>0$ large enough $(\ba)^d\cdot(\bb)$ satisfies
the claim for $Y$.

$(\ref{2independent})$. The proof is by induction on $n$. The base case is when $P_X=[n]$. In this case, let $\ba\in X$ be the tuple given by item (\ref{1independent}). It follows
immediately that the matrix $A$ with rows $(\ba)^1,(\ba)^2,\dots,(\ba)^n$ is non-singular as a Vandermonde matrix can be easily obtained if we multiply column $i$ by $1/\ba[i]$, for every $i\in[n]$. 

For the inductive case, we can assume that $P_X\neq [n]$.
Pick a vector $\by\in X$ where the set $I=I_{\by}$ defined as
$I_{\by}:=\{i\in [n] \mid \by[i]\neq 0\}$ is non-empty and as small as possible.
Note that $I\neq[n]$ since $P_X\neq[n]$. Let 
\[Y:=\{\ba\in X \mid \forall i\in I (\ba[i]\neq 0)\}\]
Note that $P_Y\subseteq I$ since $\by\in Y$ and $I\subseteq P_Y$ by the minimality of $Y$. Also, $D_Y$ contains all irreflexive pairs in $I\times I$. Indeed, for every such pair $(i,j)$, let $\bb$ be the tuple in $X$ witnessing $(i,j)$. Clearly
$\bb[i]\neq 0$ or $\bb[j]\neq 0$. It follows that $\bb\in Y$, since otherwise $I_{\by\cdot \bb}$ would 
be strictly smaller than $I$ yet non empty.

Now, let $\bz$ be the tuple obtained by 
applying item (\ref{1independent}) to $Y$.

Consider the set $Z$ defined as
\[Z:=\{\ba \in Y \mid \forall i\not\in I (\ba[i]=0)\}\]
Note that $\bz\in Z$ and $\bz[i]\neq \bz[j]$ for every $i\neq j\in I$. Then by the inductive hypothesis,
\[\pr_I Z:=\{\pr_I \ba \mid \ba\in Z\}\] 
has $|I|$ independent vectors $\pr_I \ba_1\dots,\pr_I \ba_{|I|}$.
Also, note that $D_{\pr_{\overline{I}} X}$ has all irreflexive pairs in $\overline{I}\times\overline{I}$ and,
hence, by the inductive hypothesis $\pr_{\overline{I}} X$ has $n-|I|$ independent vectors 
$\pr_{\overline{I}} \ba_{|I|+1},\dots, \pr_{\overline{I}} \ba_n$.
It follows easily that $\ba_1\dots,\ba_n$ are linearly independent.
\end{claimproof}

It will be convenient to endow a structure $\bD$ with designated elements and constraints. That is,
given $d_1,\dots,d_n$ where $d_i\in D\cup \mathcal{C}_{\bD}$ we shall use $(\bD,d_1,\dots,d_n)$ to denote the structure
obtained from $\bD$ where $d_1,\dots,d_n$ have been {\em designated}. Homomorphisms are extended in a natural way to this more general setting. That
is, a homomorphism from $(\bD,d_1,\dots,d_n)$ to $(\bE,e_1,\dots,e_n)$ is any homomorphism
$h$ from $\bD$ to $\bE$ such that, additionally, $h(d_i)=e_i$ for all $i\in[n]$ where if $d_i$ is a constraint, say,
$d_i=R(\bd)$ then $h(d_i)$ is defined to be the constraint $R(h(\bd))$. Similarly, we shall denote the number of homomorphisms from $(\bD,d_1,\dots,d_n)$ to $(\bE,e_1,\dots,e_n)$ by  $\Hom(\bD,d_1,\dots,d_n;\bE,e_1,\dots,e_n)$. A \textit{rooted ftree} is any structure $(\bT,t)$
where $\bT$ is an ftree and $t\in T$.

The following definitions, although somewhat technical, will be used a few times.
Assume that we want to compute $\Hom(\bT,t,R(\bt);\bD,d,R'(\bd))$ where $t\in T$ and $d\in D$. We say that $(t,R(\bt))$ {\em can be mapped to} $(d,R'(\bd))$ if the following conditions hold:
\begin{enumerate}
    \item $R=R'$, and
    \item for every $s,s'\in[\arity(R)]$, $\bt[s]=\bt[s']$ implies $\bd[s]=\bd[s']$, and 
    \item for every $s\in[\arity(R)]$, $\bt[s]=t$ implies $\bd[s]=d$.
\end{enumerate}
Alternatively, $(t,R(\bt))$ {\em can be mapped to} $(d,R'(\bd))$ if $R=R'$ and there is a mapping
$h:\{t\}\cup\{\bt\}\rightarrow\{d\}\cup\{\bd\}$ such that $h(t)=d$ and $h(\textbf{t})=\textbf{d}$.

Assume that, additionally, $\bT$ is an ftree.
Let $\bT_m$, $m\in M$ be the collection of all maximal subftrees of $\bT$ satisfying the following 
conditions:
\begin{enumerate}
\item $\bT_m$ does not contain $R(\bt)$ \label{Treem1}
\item Every $\bT_m$ contains an element $t_m$ in $\{\bt\}$ (which by condition (\ref{Treem1}) must be unique) and, additionally, $t_m$
participates in only one constraint in $\bT_m$.
\end{enumerate}
Alternatively, we can regard collection $\bT_m, m\in M$ as constructed in two stages from $\bT$ as follows. In a first stage, we remove constraint $R(\bt)$ from $\bT$ obtaining a collection of ftrees satisfying condition (\ref{Treem1}). Note that each one of the ftrees $\bQ$ obtained in this way contains a unique element $q$ in $\{\bt\}$
but $q$ might appear in several constraints $C_1,\dots,C_r$ of $\bQ$. Then, $\bQ$ is 
further divided in $r$ ftrees $\bQ_1,\dots,\bQ_r$ in the following way: $\bQ_i$, $i\in[r]$ is
the substructure of $\bQ$ induced 
by all the elements that are still connected to $q$ if we remove constraints $C_1,\dots,C_{i-1},C_{i+1},\dots,C_r$.

Then, we define
$\bT\setminus R(\bt)$ to be the collection $(\bT_m,t_m)$, $m\in M$ of rooted ftrees obtained in this way. The following claim is immediate.

\begin{claim} \label{cl:homproduct}
Assume that $t\in R(\bt)$. Then, 
$\Hom(\bT,t,R(\bt);\bD,d,R'(\bd))=0$ if  $(t,R(\bt))$ cannot be mapped to $(d,R'(\bd))$.
Otherwise, 
$\Hom(\bT,t,R(\bt);\bD,d,R'(\bd))=\prod_{m\in M}\Hom(\bT_m,t_m;\bD,\bd[s_m])$ where
$s_m\in[\arity(R)]$ is such that $t_m=\bt[s_m]$.
\end{claim}

Then we have the following lemmas: 
\begin{lemma}
\label{le:partitiontohom}
Let $\bD$ be a $\sigma$-structure and let $(\{P_i \mid i\in I\},\{Q_j \mid j\in J\})$ be an equitable partition
of $\bD$. Then, for every $d_1,d_2$ in the same $P$-class and every rooted ftree $(\bT,t)$, $\Hom(\bT,t;\bD,d_1)=\Hom(\bT,t;\bD,d_2)$.
\end{lemma}
\begin{proof}
The proof follows easily by induction on the number of constraints of $\bT$. Assume that
$\bT$ contains no constraints. In this case
$\bT$ consists only of node $t$ and the claim is immediate. Otherwise, let $R(\bt)$ be any constraint in $\bT$ containing $t$ and let $(\bT_m,\bt[s_m])$, $m\in M$ be $\bT\setminus R(\bt)$.

For every constraint $C\in \mathcal{C}_{\bD}$ and every $d\in D$ we shall use $n_{d,C}$ to denote
$\Hom(\bT,t,R(\bt);\bD,d,C)$. We claim that for every two constraints $C_1,C_2$ in the same $Q$-class satisfying $M_{\bD}[d_1,C_1]=M_{\bD}[d_2,C_2]$ we have $n_{d_1,C_1}=n_{d_2,C_2}$.

Let $C_1=R_1(\bd_1)$ and $C_2=R_2(\bd_2)$. Note that since $M_{\bD}[d_1,C_1]=M_{\bD}[d_2,C_2]$
we can assume that $C_1$ and $C_2$ share the same relation symbol. Let us do a case analysis using Claim \ref{cl:homproduct}:

\begin{itemize}
    \item If $(t,R(\bt))$ does not map to $(d_1,R_1(\bd_1))$ then $n_{d_1,C_1}=0$. Since $C_2$ belongs to the same $Q$-class and $M_{\bD}[d_1,C_1]=M_{\bD}[d_2,C_2]$ it follows
    that $(t,R(\bt))$ does not map to $(d_2,R_2(\bd_2))$ either, and hence $n_{d_2,C_2}=0$.
    \item Otherwise  $n_{d_q,C_q}=\prod_{m\in M}\Hom(\bT_m,t_m;\bD,\bd_q[s_m])$ ($q=1,2$). Since $C_1$ and $C_2$ belong to the same $Q$-class, it follows that $\bd_1[s]$ and $\bd_2[s]$ belong to the same $P$-class for every $s\in[\arity(R)]$. It then follows by induction
      that $n_{d_1,C_1}=n_{d_2,C_2}$.
\end{itemize}

Then, for every $i \in I$, $j\in J$, and $\ell\in L$ there exists some unique $n^\ell_{i,j}$ such that $n_{d,C}=n^\ell_{i,j}$ for every $d\in P_i$ and every $C\in Q_j$ with $M_{\bD}[d,C]=\ell$.
Let $P_i$ be the $P$-class of $d_1$ and $d_2$. To finalize the proof it is enough to see that for every $q\in\{1,2\}$, \[\Hom(\bT,t;\bD,d_q)=\sum_{C\in \mathcal{C}_{\bD}} \Hom(\bT,t,R(\bt);\bD,d_q,C)=\sum_{C\in \mathcal{C}_{\bD}} n_{d_q,C}=
\sum_{j\in J,\ell\in L} n^\ell_{i,j} c^\ell_{i,j}\]
where $c^\ell_{i,j}$ is as in condition (\ref{eq:eqpartQ}) of equitable partition. Hence, in the previous expression we use that $c^\ell_{i,j}=|\{C\in Q_j \mid M_{\bD}[d_q,C]=\ell\}|$ for $q \in \{1,2\}$.

\end{proof}

\begin{lemma}
\label{le:homtopartition}
Let $\bD$ be a $\sigma$-structure, let $(\{P_i \mid i\in I\},\{Q_j \mid j\in J\})$ defined as follows:
\begin{enumerate}
    \item $P$ is the partition of $D$ that places two elements $d_1,d_2$ in the same class if
    $\Hom(\bT,t;\bD,d_1)=\Hom(\bT,t;\bD,d_2)$ for every rooted ftree $(\bT,t)$;
    \item $Q$ is the partition of $\mathcal{C}_{\bD}$ that places two constraints $R_1(\bd_1)$, $R_2(\bd_2)$ in the same class if the following conditions hold:
    \begin{itemize}
    \item $R_1=R_2$, and
    \item for every $s,s'\in[\arity(R)]$, $\bd_1[s]=\bd_1[s']$ iff $\bd_2[s]=\bd_2[s']$, and
    \item for every $s\in[\arity(R)]$, $\bd_1[s]$ and $\bd_2[s]$ belong to the same $P$-class.
    \end{itemize}
\end{enumerate}
Then $(P,Q)$ is an equitable partition.
\end{lemma}

\begin{proof}

We only need to prove that for every $i\in I$, every $j\in J$,
and every $\ell\in L$ there exists $c^\ell_{i,j}$ such that 
$c^{\ell}_{d,j}=c^{\ell}_{i,j}$
for every $d\in P_i$, where $c^{\ell}_{d,j}=|\{C\in Q_j \mid M_{\bD}[d,C]=\ell\}|$
(since the other condition of equitable partition follows immediately from the
definition of $(P,Q)$). 

Let ${\mathcal Z}$ be the set of all $(\bT,t,R(\bt))$ where $\bT$ is an ftree, $R(\bt)\in \mathcal{C}_{\bT}$, and $t\in\{\bt\}$.

We shall start by proving that for every $i\in I$, $j\in J$, $\ell\in L$, and $z \in {\mathcal Z}$, there exists 
$n^{z,\ell}_{i,j}$ such that $n^{z}_{d,C}=n^{z,\ell}_{i,j}$ for every $d\in P_i$ and every $C\in Q_j$, where $n^z_{d,C}$ is defined to be  $\Hom(\bT,t,R(\bt);\bD,d,C)$. That is, we want to prove the following claim.

\begin{claim}
For every $d_1,d_2$ inside the same class $P_i$ and every $C_1,C_2$ inside the same class $Q_j$ such that $M_{\bD}[d_1,C_1]=M_{\bD}[d_2,C_2]$ we have $n^{z}_{d_1,C_1}=n^{z}_{d_2,C_2}$. 
\end{claim}
\begin{claimproof}{6}
Again, let us do a case analysis using Claim \ref{cl:homproduct}:

\begin{itemize}
\item If $(t,R(\bt))$ cannot be mapped to $(d_1,C_1)$ then since $C_2$ belongs to the same $Q$-class as $C_1$ and $M_{\bD}[d_1,C_1]=M_{\bD}[d_2,C_2]$
it follows that $(t,R(\bt))$ cannot be mapped to $(d_2,C_2)$ either. Then,
we have $n^{z}_{d_1,C_1}=n^{z}_{d_2,C_2}=0$.
\item Assume that we are not in the previous case. Let $C_1=R(\bd_1)$ and $C_2=R(\bd_2)$
and let $(\bT_m,\bt[s_m])$, $m\in M$ be $\bT\setminus R(\bt)$. Then,
we have that for every $q\in \{1,2\}$ $n^{z}_{d_q,C_q}=\prod_{m\in M}\Hom(\bT_m,\bt[s_m];\bD,\bd_q[s_m])$. Since $C_1$ and $C_2$ belong to the same $Q$-class, it follows that $\bd_1[s]$ and $\bd_2[s]$ belong to the same $P$-class for every $s\in[\arity(R)]$. It then follows from the definition of $(P,Q)$ that $n^{z}_{d_1,C_1}=n^{z}_{d_2,C_2}$.\end{itemize}
\end{claimproof}

Then for every $i\in I$ and every $d\in P_i$ we have
\begin{equation}\label{eq:hom}\Hom(\bT,t;\bD,d)=\sum_{C\in \mathcal{C}_{\bD}} \Hom(\bT,t,R(\bt);\bD,d,C)=\sum_{j\in J,\ell\in L} n^{z,\ell}_{i,j} c^{\ell}_{d,j}=\sum_{(j,\ell)\in K_i} 
n^{z,\ell}_{i,j} c^{\ell}_{d,j}\end{equation}

where $K_i=\{(j,\ell)\in J\times L \mid n^{z,\ell}_{i,j}\neq 0 \text{ for some }z\in {\mathcal Z}\}$. The following is immediate.

\begin{claim}
\label{cl:existsC}
Let $i\in I$ and $d\in P_i$. For every $(j,\ell)\in J\times L$, $(j,\ell)\in K_i$ iff there exists a constraint $C=R'(\bd)$ in $Q_j$ with $M_{\bD}[d,C]=\ell$ and $d\in\{\bd\}$.
\end{claim}

In what remains of the proof $i$ is any arbitrary element of $I$. Then, it follows from Claim \ref{cl:existsC} that if $(j,\ell)\not\in K_i$ we can safely set $c^{\ell}_{i,j}=0$. 
It now remains to show the existence of $c^{\ell}_{i,j}$ for every $(j,\ell)\in K_i$. 
To this end we only need to
observe that (\ref{eq:hom}) holds for every $z\in {\mathcal Z}$ and invoke Claim \ref{cl:independent}(\ref{2independent}) to conclude that 
$c^{d_1,\ell}_{i,j}=c^{d_2,\ell}_{i,j}$ for every $j\in J$,$\ell\in L$
and every $d_1,d_2\in P_i$.

However, we need first to guarantee that the hypothesis of the claim are met.
First, we prove the following:
\begin{claim}
Let $(j_1,\ell_1),(j_2,\ell_2)$ be different elements in $K_i$. Then,  
there exists $z\in{\mathcal Z}$
such that $n^{z,\ell_1}_{i,j_1}\neq n^{z,\ell_2}_{i,j_2}$.
\end{claim}
\begin{claimproof}{8}
Let $d\in P_i$, and let $C_q=R_q(\bd_q), q\in\{1,2\}$ be a constraint in $Q_{j_q}$ with $M_{\bD}[d,C_q]=\ell_q$
such that $d\in\{\bd_q\}$. If $(d,C_1)$ cannot be
 mapped to $(d,C_2)$ then we just need to set $z$ to be $(\bT,d,C_1)$ where $\bT$ 
is the ftree with universe $\{\bd_1\}$ containing only constraint $C_1$. A similar construction takes care of the case when $(d,C_2)$ cannot 
be mapped to $(d,C_1)$. Hence, we are left with the case in which each of $(d,C_1),(d,C_2)$ can be mapped to each other, that is, where $(d,C_1),(d,C_2)$ are identical modulo renaming the elements. This implies that $\ell_1=\ell_2$ and, hence,  $j_1\neq j_2$, which implies that that there exists $s\in[\arity(R_1)]$ such that $\bd_1[s]$
and $\bd_2[s]$ belong to a different $P$-class. Consequently, by the definition of $(P,Q)$ we have that $\Hom(\bT,t;\bD,\bd_1[s])\neq
\Hom(\bT,t;\bD,\bd_2[s])$ for some rooted ftree $(\bT,t)$. Now, let $z=(\bQ,d,R_1(\bd_1))$
where $\bQ$ is the ftree
obtained by adding constraint $R_1(\bd_1)$ to $\bT$ (we can assume, renaming variables
if necessary that $T$ does not contain any element in $\{\bd_1\}$) and identifying
$t$ with $\bd_1[s]$. It follows immediately
that $n^{z,\ell_q}_{i,j_q}=\Hom(\bQ,d,R_1(\bd_1);\bD,d,R_q(\bd_q))=\Hom(\bT,t;\bD,\bd_q[s])$ and we are done.
\end{claimproof}

Denote by ${\bf n}^{z}$ the $K_i$-vector with entries $n^{z,\ell}_{i,j}$, $(j,\ell) \in K_i$.

\begin{claim}
Let $z_q=(\bT_{q},t_q,R_q(\bt_{q}))\in{\mathcal Z}$, $q=1,2$ such that ${\bf n}^{z_1}\cdot {\bf n}^{z_2}\neq {\bf 0}$.
Then there exists $z\in {\mathcal Z}$ such that ${\bf n}^z={\bf n}^{z_1}\cdot {\bf n}^{z_2}$
\end{claim}
\begin{claimproof}{9}
Clearly if $R_1\neq R_2$ then $n^{z_1,\ell}_{i,j}\cdot n^{z_2,\ell}_{i,j}=0$ for every $(j,\ell)\in K_i$ and nothing needs to be done. 
Otherwise, let $z=(\bT,t,R(\bt))$ where $\bT$ is obtained by first computing the disjoint union of $\bT_1$ and $\bT_2$ and then, for every 
$s\in[\arity(R)]$, merging $\bt_1[s]$ and $\bt_2[s]$ into a single node, as well as  identifying $t_1$ and $t_2$ into a single element $t$. Note that after the identification $R_1(\bt_1)$ and 
$R_2(\bt_2)$ become the same constraint, which we denote by $R(\bt)$.

It only remains to see that $n^{z_1,\ell}_{i,j}\cdot n^{z_2,\ell}_{i,j}=n^{z,\ell}_{i,j}$ for every $(j,\ell)\in K_i$. Let $d\in P_i$, and let $C\in Q_j$ be such
that $\ell=M_{\bD}[d,C]$. Clearly, if $(t_1,R(\bt_1))$
cannot be mapped to $(d,C)$ then $n^{z_1,\ell}_{i,j}=0$. Note that in this case 
$(t,R(\bt))$ cannot be mapped to $(d,C)$ either, and hence $n^{z,\ell}_{i,j}=0$ as desired. A
similar reasoning applies if $(t_2,R(\bt_2))$ cannot be mapped to $(d,C)$ and, hence,
we only need to deal with the case in which both $(t_1,R(\bt_1))$ and $(t_2,R(\bt_2))$
can be mapped to $(d,C)$. In this case it follows that $(t,R(\bt))$ can be mapped to $(d,C)$. 
Let $(\bT_m,\bt[s_m])$, $m\in M$ be $\bT\setminus R(\bt)$ and let $C=R(\bd)$.
Then
\[n^{z}_{d,C}=\Hom(\bT,t,R(\bt);\bD,d,R(\bd))=\prod_{m\in M}\Hom(\bT_m,\bt[s_m];\bD,\bd[s_m])\]
Note that $M$ can be partitioned in two sets $M_1,M_2$ such that for every $q\in \{1,2\}$
$\bT_q \setminus R(\bt_q)$ is precisely $(\bT_m,\bt[s_m])$, $m\in M_q$.

It follows that 
\begin{align*}n^{z,\ell}_{i,j} = n^{z}_{d,C} &=\prod_{m\in M}\Hom(\bT_m,\bt[s_m];\bD,\bd[s_m]) \\
&=\prod_{m\in M_1}\Hom(\bT_m,\bt[s_m];\bD,\bd[s_m]\cdot \prod_{m\in M_2}\Hom(\bT_m,\bt[s_m];\bD,\bd[s_m]) \\
&=\Hom(\bT_1,t_1,R(\bt_1);\bD,d,R(\bd))\cdot \Hom(\bT_2,t_2,R(\bt_2);\bD,d,R(\bd)) \\
&=n^{z_1}_{d,C}\cdot n^{z_2}_{d,C}=n^{z_1,\ell}_{i,j}\cdot n^{z_2,\ell}_{i,j}
\end{align*}
as desired.
\end{claimproof}

\end{proof}

We are finally ready to give a proof of the equivalence of $(\ref{3digraphItem})$ and $(\ref{4digraphItem})$.

$(\ref{3digraphItem})\Rightarrow (\ref{4digraphItem})$. 
 Let $\bD$ denote the disjoint union of $\bA$ and $\bB$, 
let $(\{P_i \mid i\in I\},\{Q_j \mid j\in J\})$ be an equitable partition of $\bD$ witnessing that $\bA$ and $\bB$ have a common equitable partition, and let $\bT$ be an ftree and $t\in T$.
It follows from Lemma \ref{le:partitiontohom} that for every $i\in I$ there is a value $n_i$ such that
for every $d\in P_i$, $\Hom(\bT,t;\bD,d)=n_i$. 

Then, for every
$\bE\in \{\bA,\bB\}$, \[\Hom(\bT,\bE)=\sum_{e\in E} \Hom(\bT,t;\bE,e)=\sum_{i\in I} n_i \cdot p^{\bE}_i\]
where $p^{\bE}_i=|P_i\cap E|$. Since $(P,Q)$ is an equitable partition of $\bA \cup \bB$, it follows that $p^{\bA}_i=p^{\bB}_i$ for every $i\in I$ and we are done. 

$(\ref{4digraphItem})\Rightarrow (\ref{3digraphItem})$. Let $\bD$  be the disjoint union of $\bA$ and $\bB$ and let $(\{P_i \mid i\in I\},\{Q_j \mid j\in J \})$ be the partition of $\bD$ defined as in Lemma \ref{le:homtopartition}. We first show that $p^{\bA}_i=p^{\bB}_i$ for every $i\in I$, where $p^{\bE}_i=|P_i\cap E|$ for $\bE \in \{\bA,\bB\}$.

Let $z=(\bT,t)$ be any rooted ftree. Note that from the definition of $(P,Q)$
it follows that for every $i\in I$ there exists $n^z_i$ such that
$n^z_i=\Hom(\bT,t;\bD,d)$ for every $d\in P_i$. Consequently, for $\bE \in \{\bA,\bB\}$ we have:

\[\Hom(\bT,\bE)=\sum_{e\in E}\Hom(\bT,t;\bE,e)=\sum_{i\in I} n^z_i \cdot p^{\bE}_i\]

Since the previous identity holds for every rooted ftree $z$, it is only necessary to invoke Claim \ref{cl:independent} to conclude 
that $p^{\bA}_i=p^{\bB}_i$ for every $i\in I$. However, we must first verify
that the conditions of Claim \ref{cl:independent} are satisfied.

First, we need to show that for every $i\neq i'\in I$, there exists a rooted ftree $z$ such that $n^z_i\neq n^z_{i'}$. This follows immediately from the definition of $P$. Secondly,
we need to show that for every pair of rooted ftrees $z_1=(\bT_1,t_1)$, $z_2=(\bT_2,t_2)$, there exists a rooted ftree $z=(\bT,t)$ such that $n^z_i=n^{z_1}_i\cdot n^{z_2}_i$. Indeed,
it is easy to verify that the condition is satisfied if we let $\bT$ be the rooted ftree obtained by merging $t_1$ and $t_2$ into a single node $t$ in the disjoint union of $\bT_1$ and $\bT_2$.

\section{Proof of Lemma \ref{le:homTreewidth}}

\homTreewidth*

From Theorem \ref{le:digraph} it follows that for a pair of $\sigma$-structures $\bfa$ and $\bfb$, $\astarkbf \eqwl[k] \bstarkbf$ if and only if $\Hom(\bT,\astarkbf)=\Hom(\bT,\bstarkbf)$ for every $\sigma^*_k$-ftree $\bT$. So it only remains to prove the following.

\begin{claim}
Assume that $r\leq k$. Then the following are equivalent:
\begin{enumerate}
\item $\Hom(\bQ,\bA)=\Hom(\bQ,\bB)$ for every $\sigma$-structure $\bQ$ of treewidth $<k$; \label{1treewidth}
\item $\Hom(\bT,\astarkbf)=\Hom(\bT,\bstarkbf)$ for every $\sigma^*_k$-ftree $\bT$. \label{2treewidth}
\end{enumerate}
\end{claim}

\begin{claimproof}{10}

$(\ref{1treewidth})\Rightarrow (\ref{2treewidth})$.
Let $\bT$ be a $\sigma^*$-ftree. It follows immediately that if (\ref{1treewidth}) holds then both $\bA$ and $\bB$ must have the same number of elements and constraints. It then follows that (\ref{2treewidth}) holds for $\bT$ if it consists of a single element and no constraints at all. Consequently we can safely assume that all elements in $\bT$ participate in at least one constraint.

In what follows $\dstarkbf\in\{\astarkbf,\bstarkbf\}$.
Let $t$ be any node in $T$. Since $t$ participates in a constraint it follows that the possible image of $t$ in  homomorphism from $\bT$
is heavily restricted. In particular, if the image of $t$ according to some homomorphism from $\bT$ to $\dstarkbf$ is in $D^j$ for some $j\leq k$ then necessarily the image of $t$ in any homomorphism from $\bT$ to any structure $\cstarkbf\in \{\astarkbf,\bstarkbf\} $ must be in $C^j$. This means that we can safely add constraint
$T_{j,\emptyset}(t)$ to $\bT$ without altering $\Hom(\bT,\astarkbf)$ or 
$\Hom(\bT,\bstarkbf)$.

Likewise, if some homomorphism from $\bT$ to a structure $\dstarkbf$ maps $t$ to a constraint $R(\bd)$, then likewise we can assume that constraint $R_{\emptyset}(t)$ belongs to $\bT$.

To complete the proof we shall show that it is always possible to construct from $\bT$ a $\sigma$-structure $\bQ$ of treewidth $<k$ such that $|\Hom(\bT,\dstarkbf)|=|\Hom(\bQ,\bD)|$. It is convenient
to construct $\bQ$ in two stages. First, let us construct a $\sigma$-structure $\bP$ (not necessarily of treewidth $<k$) satisfying that $|\Hom(\bT,\dstarkbf)|=|\Hom(\bP,\bD)|$. We shall allow to use
equalities in $\bP$, i.e., constraints of the form $p_1=p_2$, indicating that $p_1$ and $p_2$ must be assigned to the same element in $D$. 

We shall define $\bP$ along with a function $\alpha$ mapping every element $t$ of $\bT$ to a $j$-ary tuple 
of elements in $\bP$ ($j\leq k$) inductively on the number of elements of $\bT$ as follows.

Assume (base case) that $\bT$ contains a unique element $t$. As discussed above we can assume that $\bT$ contains
constraint $T_{j,\emptyset}(t)$ for some $j\leq k$ or $R_{\emptyset}(t)$ for some $R\in\sigma$. In the first
case, we set the universe of $\bP$ to contain $j$ new elements $p_1,\dots,p_j$. Furthermore, for every
unary constraint $T_{j,S}(t)$ in $\bT$ and every $i,i'\in S$, we include in $\bP$ the equality $p_i=p_{i'}$, and we define $\alpha(t)=(p_1,\ldots,p_j)$.
In the second case, we set the universe of $\bP$ to contain $\arity(R)$ new elements $p_1\dots,p_{\arity(R)}$ and we include in $\bP$ the constraint 
$R(p_1,...,p_{\arity(R)})$. Similarly to the previous case,
for every unary constraint $R_{S}(t)$ in $\bT$ and every $i,i'\in S$, we include in $\bP$ the equality $p_i=p_{i'}$. Finally, we set $\alpha(t)=(p_1,\ldots,p_{\arity(R)})$.

Let us consider now the inductive case. Let $t_1$ and $t_2$ be nodes that participate in a binary constraint $U(t_1,t_2)$ (recall that $U$ is either $\tjii$ or $R_{\bf i}$) in $\bT$. By removing this constraint $\bT$ gets divided in two ftrees $\bT_1$ and $\bT_2$ such that $\bT_1$ contains $t_1$ and $\bT_2$ contains $t_2$. Now, assume that $\bP_i$ and $\alpha_i$ are already constructed for $\bT_i$, $i=1,2$. We are ready to define $\bP$. First, we  compute the disjoint union of $\bP_1$ and $\bP_2$. Then, we add some further equalities depending on constraint $U(t_1,t_2)$. Consider first the case that $U=T_{j_1,{\bf i}}$, ${\bf i}=(i_1,\dots,i_{j_2})\in [j_1]^{j_2}$ for some $j_1,j_2\leq k$ and let
$\alpha_i(t_i)=(p^i_1,\dots,p^i_{j_i})$, $i=1,2$. Then, for every $\ell\leq j_2$ we add the equality $p^2_\ell=p^1_{i_{\ell}}$.
Finally, for every $t\in T$ we define $\alpha(t)$ to be $\alpha_i(t)$ where $\bT_i$ contains $t$. The procedure is identical for $U=\rii$, where we just substitute $j_1$ by $\arity(R)$. It follows immediately from the definition
that $\Hom(\bP,\bD)=\Hom(\bT,\dstarkbf)$.

Finally, let us define $\bQ$ to be the structure obtained by identifying (i.e, merging into a single element) all elements in $\bP$ joined by a chain of equalities. It is immediate that $\Hom(\bP,\bD)=\Hom(\bQ,\bD)$.

We shall conclude by giving a tree-decomposition $(G,\beta)$ of $\bQ$ of width $<k$. In particular, let $G$ be the tree where the vertex set is precisely the universe of $\bT$ and two different nodes are adjacent if both participate in some common constraint in $\bT$ and let $\beta(t)=\{\alpha(t)\}$.

$(\ref{2treewidth})\Rightarrow (\ref{1treewidth})$. Let $\bQ$ be a $\sigma$-structure of treewidth $<k$ and let $\bD\in\{\bA,\bB\}$. We note here
that we can assume that $\bQ$ is connected since if $\bQ$ is the disjoint union
of structures $\bQ_1$ and $\bQ_2$ then $\Hom(\bQ,\bD)=\Hom(\bQ_1,\bD)\cdot \Hom(\bQ_2,\bD)$.
We shall show that there exists a $\sigma_k^*$-ftree
$\bT$ such that $\Hom(\bT,\dstarkbf)=\Hom(\bQ,\bD)$. 
Let $(G,\beta)$ be a tree-decomposition
of width at most $k$ of $\bQ$. It is well-known and easy to prove that, since $\bQ$ is connected we can always construct $(G,\beta)$ in such a way
that for every pair $u,v$ of adjacent nodes in $G$, $\beta(u)\subseteq\beta(v)$ or $\beta(v)\subseteq\beta(u)$. Furthermore,
it is easy to see that we can further enforce that for every constraint $R(\bq)$ in $\bQ$ there exists a node $v\in G$ such that
$\beta(v)=\{\bq\}$.

The universe $T$ of $\bT$ is $V\cup \mathcal{C}_{\bQ}$ where $V$ is the node-set of $G$. Furthermore $\bT$ contains the following constraints.

Let us start with the unary constraints. Let $t$ be an element in $\bT$. If $t=v\in G$ then we include
in $\bT$ a constraint $T_{j_v,\emptyset}(t)$ where $j_v=|\beta(v)|$. Otherwise, if $t=R(\bq)\in \mathcal{C}_{\bQ}$ then we include
in $\bT$ all constraints $R_{\{i,i'\}}(t)$ where $\bq[i]=\bq[i']$.

Now, let us turn our attention to the binary constraints. Fix some arbitrary ordering on $Q$ and for every $v\in V$
let $\bq^v=(q^v_1,\dots,q^v_{j_v})$ be an array containing the nodes in $\beta(v)$ following this fixed order. 

Then, for every edge $(u,v)$ in $G$ include constraint $T_{j_u,{\bf i}}(u,v)$ where ${\bf i}=(i_1,\dots,i_{j_v})$ is defined as follows. First, we assume without loss of generality that $\beta(v)\subseteq\beta(u)$. Then, for every $\ell\leq j_v$, $i_{\ell}$ is defined to
be such that $\bq^u[i_{\ell}]=\bq^v[\ell]$.

Finally, for every constraint $t=R(\bq)$ in $\bQ$ we pick some element $v\in V$ satisfying $\{\bq\}=\beta(v)$ and
we add the constraint $R_{\bf i}(t,v)$ with ${\bf i}=(i_1,\dots,i_{j_v})$ where $i_{\ell}$ satisfies $\bq[i_{\ell}]=\bq^v[\ell]$. It is immediate to see that $\bT$ is an ftree and that $\Hom(\bQ,\bD)=\Hom(\bT,\dstarkbf)$\end{claimproof}

%% file: ms.bbl
\begin{thebibliography}{10}

\bibitem{Angluin1980Local}
Dana Angluin.
\newblock Local and global properties in networks of processors (extended
  abstract).
\newblock In {\em Proceedings of the Twelfth Annual ACM Symposium on Theory of
  Computing}, STOC ’80, page 82–93, New York, NY, USA, 1980. Association
  for Computing Machinery.

\bibitem{AtseriasBD09}
Albert Atserias, Andrei~A. Bulatov, and Anuj Dawar.
\newblock Affine systems of equations and counting infinitary logic.
\newblock {\em Theor. Comput. Sci.}, 410(18):1666--1683, 2009.

\bibitem{Atserias2013}
Albert Atserias and Elitza Maneva.
\newblock Sherali--adams relaxations and indistinguishability in counting
  logics.
\newblock {\em SIAM Journal on Computing}, 42(1):112--137, 2013.

\bibitem{Babai2015}
L\'{a}szl\'{o} Babai.
\newblock Graph isomorphism in quasipolynomial time [extended abstract].
\newblock In {\em Proceedings of the Forty-Eighth Annual ACM Symposium on
  Theory of Computing}, STOC '16, page 684–697, New York, NY, USA, 2016.
  Association for Computing Machinery.

\bibitem{Babai1980}
L{\'{a}}szl{\'{o}} Babai, Paul Erd{\"{o}}s, and Stanley~M. Selkow.
\newblock Random graph isomorphism.
\newblock {\em {SIAM} J. Comput.}, 9(3):628--635, 1980.

\bibitem{Barto2016Collapse}
Libor Barto.
\newblock The collapse of the bounded width hierarchy.
\newblock {\em Journal of Logic and Computation}, 26(3):923--943, 2016.

\bibitem{BrakensiekGWZ20}
Joshua Brakensiek, Venkatesan Guruswami, Marcin Wrochna, and Stanislav
  Zivn{\'{y}}.
\newblock The power of the combined basic linear programming and affine
  relaxation for promise constraint satisfaction problems.
\newblock {\em {SIAM} J. Comput.}, 49(6):1232--1248, 2020.

\bibitem{Bulatov2017}
A.~A. {Bulatov}.
\newblock A dichotomy theorem for nonuniform {CSPs}.
\newblock In {\em 2017 IEEE 58th Annual Symposium on Foundations of Computer
  Science (FOCS)}, pages 319--330, Oct 2017.

\bibitem{Butti2021}
Silvia Butti and Victor Dalmau.
\newblock The complexity of the distributed constraint satisfaction problem.
\newblock In Markus Bl{\"{a}}ser and Benjamin Monmege, editors, {\em 38th
  International Symposium on Theoretical Aspects of Computer Science, {STACS}
  2021, March 16-19, 2021, Saarbr{\"{u}}cken, Germany (Virtual Conference)},
  volume 187 of {\em LIPIcs}, pages 20:1--20:18. Schloss Dagstuhl -
  Leibniz-Zentrum f{\"{u}}r Informatik, 2021.

\bibitem{Cai92}
Jin{-}yi Cai, Martin F{\"{u}}rer, and Neil Immerman.
\newblock An optimal lower bound on the number of variables for graph
  identifications.
\newblock {\em Comb.}, 12(4):389--410, 1992.

\bibitem{ChanLRS13}
Siu~On Chan, James~R. Lee, Prasad Raghavendra, and David Steurer.
\newblock Approximate constraint satisfaction requires large {LP} relaxations.
\newblock In {\em 54th Annual {IEEE} Symposium on Foundations of Computer
  Science, {FOCS} 2013, 26-29 October, 2013, Berkeley, CA, {USA}}, pages
  350--359. {IEEE} Computer Society, 2013.

\bibitem{DalmauK13}
V{\'{\i}}ctor Dalmau and Andrei~A. Krokhin.
\newblock Robust satisfiability for csps: Hardness and algorithmic results.
\newblock {\em {ACM} Trans. Comput. Theory}, 5(4):15:1--15:25, 2013.

\bibitem{DalmauKM18}
V{\'{\i}}ctor Dalmau, Andrei~A. Krokhin, and Rajsekar Manokaran.
\newblock Towards a characterization of constant-factor approximable
  finite-valued {CSPs}.
\newblock {\em J. Comput. Syst. Sci.}, 97:14--27, 2018.

\bibitem{Dell2018}
Holger Dell, Martin Grohe, and Gaurav Rattan.
\newblock Lov{\'{a}}sz meets weisfeiler and leman.
\newblock In Ioannis Chatzigiannakis, Christos Kaklamanis, D{\'{a}}niel Marx,
  and Donald Sannella, editors, {\em 45th International Colloquium on Automata,
  Languages, and Programming, {ICALP} 2018, July 9-13, 2018, Prague, Czech
  Republic}, volume 107 of {\em LIPIcs}, pages 40:1--40:14. Schloss Dagstuhl -
  Leibniz-Zentrum f{\"{u}}r Informatik, 2018.

\bibitem{feder1998computational}
Tom{\'a}s Feder and Moshe~Y Vardi.
\newblock The computational structure of monotone monadic snp and constraint
  satisfaction: A study through datalog and group theory.
\newblock {\em SIAM Journal on Computing}, 28(1):57--104, 1998.

\bibitem{Fioretto2018}
Ferdinando Fioretto, Enrico Pontelli, and William Yeoh.
\newblock Distributed constraint optimization problems and applications: A
  survey.
\newblock {\em J. Artif. Int. Res.}, 61(1):623--698, January 2018.

\bibitem{GeorgiouMT09}
Konstantinos Georgiou, Avner Magen, and Madhur Tulsiani.
\newblock Optimal sherali-adams gaps from pairwise independence.
\newblock In Irit Dinur, Klaus Jansen, Joseph Naor, and Jos{\'{e}} D.~P. Rolim,
  editors, {\em Approximation, Randomization, and Combinatorial Optimization.
  Algorithms and Techniques, 12th International Workshop, {APPROX} 2009, and
  13th International Workshop, {RANDOM} 2009, Berkeley, CA, USA, August 21-23,
  2009. Proceedings}, volume 5687 of {\em Lecture Notes in Computer Science},
  pages 125--139. Springer, 2009.

\bibitem{GhoshT18}
Mrinalkanti Ghosh and Madhur Tulsiani.
\newblock From weak to strong linear programming gaps for all constraint
  satisfaction problems.
\newblock {\em Theory Comput.}, 14(1):1--33, 2018.

\bibitem{GroheKMS14}
Martin Grohe, Kristian Kersting, Martin Mladenov, and Erkal Selman.
\newblock Dimension reduction via colour refinement.
\newblock In Andreas~S. Schulz and Dorothea Wagner, editors, {\em Algorithms -
  {ESA} 2014 - 22th Annual European Symposium, Wroclaw, Poland, September 8-10,
  2014. Proceedings}, volume 8737 of {\em Lecture Notes in Computer Science},
  pages 505--516. Springer, 2014.

\bibitem{Grohe2015Pebble}
Martin Grohe and Martin Otto.
\newblock Pebble games and linear equations.
\newblock {\em J. Symb. Log.}, 80(3):797--844, 2015.

\bibitem{HellNesetril:book}
Pavol Hell and Jaroslav Nešetřil.
\newblock {\em Graphs and Homomorphisms}.
\newblock Oxford University Press, 2004.

\bibitem{ImmermanL90}
Neil Immerman and Eric Lander.
\newblock Describing graphs: A first-order approach to graph canonization.
\newblock In {\em Complexity Theory Retrospective: In Honor of Juris Hartmanis
  on the Occasion of His Sixtieth Birthday}, pages 59--81, 1990.

\bibitem{KumarMTV11}
Amit Kumar, Rajsekar Manokaran, Madhur Tulsiani, and Nisheeth~K. Vishnoi.
\newblock On lp-based approximability for strict csps.
\newblock In Dana Randall, editor, {\em Proceedings of the Twenty-Second Annual
  {ACM-SIAM} Symposium on Discrete Algorithms, {SODA} 2011, San Francisco,
  California, USA, January 23-25, 2011}, pages 1560--1573. {SIAM}, 2011.

\bibitem{Kun2012}
Gabor Kun, Ryan O’Donnell, Suguru Tamaki, Yuichi Yoshida, and Yuan Zhou.
\newblock Linear programming, width-1 {CSPs}, and robust satisfaction.
\newblock In {\em Proceedings of the 3rd Innovations in Theoretical Computer
  Science Conference}, ITCS ’12, page 484–495, New York, NY, USA, 2012.
  Association for Computing Machinery.

\bibitem{LaroseLT07}
Beno{\^{\i}}t Larose, Cynthia Loten, and Claude Tardif.
\newblock A characterisation of first-order constraint satisfaction problems.
\newblock {\em Log. Methods Comput. Sci.}, 3(4), 2007.

\bibitem{leman1968reduction}
AA~Leman and B~Weisfeiler.
\newblock A reduction of a graph to a canonical form and an algebra arising
  during this reduction.
\newblock {\em Nauchno-Technicheskaya Informatsiya}, 2(9):12--16, 1968.

\bibitem{Lovasz1991}
L{\'{a}}szl{\'{o}} Lov{\'{a}}sz and Alexander Schrijver.
\newblock Cones of matrices and set-functions and 0-1 optimization.
\newblock {\em {SIAM} J. Optim.}, 1(2):166--190, 1991.

\bibitem{Malkin2014}
Peter Malkin.
\newblock Sherali–adams relaxations of graph isomorphism polytopes.
\newblock {\em Discrete Optimization}, 12:73--97, 2014.

\bibitem{Nesestril1989}
Jaroslav Nešetřil and Vojtěch Rödl.
\newblock Chromatically optimal rigid graphs.
\newblock {\em Journal of Combinatorial Theory, Series B}, 46(2):133--141,
  1989.

\bibitem{Ramana1994}
Motakuri~V. Ramana, Edward~R. Scheinerman, and Daniel Ullman.
\newblock Fractional isomorphism of graphs.
\newblock {\em Discrete Mathematics}, 132(1):247 -- 265, 1994.

\bibitem{Scheinerman2011fractional}
Edward~R Scheinerman and Daniel~H Ullman.
\newblock {\em Fractional graph theory: a rational approach to the theory of
  graphs}.
\newblock Courier Corporation, 2011.

\bibitem{Sherali1990}
Hanif~D. Sherali and Warren~P. Adams.
\newblock A hierarchy of relaxations between the continuous and convex hull
  representations for zero-one programming problems.
\newblock {\em {SIAM} J. Discret. Math.}, 3(3):411--430, 1990.

\bibitem{ThapperZ17}
Johan Thapper and Stanislav Zivn{\'{y}}.
\newblock The power of sherali-adams relaxations for general-valued csps.
\newblock {\em {SIAM} J. Comput.}, 46(4):1241--1279, 2017.

\bibitem{Tinhofer1986}
Gottfried Tinhofer.
\newblock Graph isomorphism and theorems of birkhoff type.
\newblock {\em Computing}, 36(4):285--300, 1986.

\bibitem{Tinhofer1991}
Gottfried Tinhofer.
\newblock A note on compact graphs.
\newblock {\em Discret. Appl. Math.}, 30(2-3):253--264, 1991.

\bibitem{YoshidaZ14}
Yuichi Yoshida and Yuan Zhou.
\newblock Approximation schemes via sherali-adams hierarchy for dense
  constraint satisfaction problems and assignment problems.
\newblock In Moni Naor, editor, {\em Innovations in Theoretical Computer
  Science, ITCS'14, Princeton, NJ, USA, January 12-14, 2014}, pages 423--438.
  {ACM}, 2014.

\bibitem{Zhuk2017}
Dmitriy {Zhuk}.
\newblock A proof of {CSP} dichotomy conjecture.
\newblock In {\em 2017 IEEE 58th Annual Symposium on Foundations of Computer
  Science (FOCS)}, pages 331--342, Oct 2017.

\end{thebibliography}
